\documentclass[pre,twocolumn,amsmath,amssymb,floatfix,superscriptaddress]{revtex4-1}

\usepackage{graphicx}
\usepackage{dcolumn}
\usepackage{bm}

\usepackage{subfigure}

\graphicspath{{images/}{plots/}}

\newcommand{\avg}[1]{\ensuremath{\left< #1 \right>}}

\newcommand{\abs}[1]{\ensuremath{\left\vert#1\right\vert}}
\newcommand{\brac}[1]{\ensuremath{\left(#1\right)}}

\begin{document}

    \title{Large deviations of connected components in the stochastic block model}
    \author{Hendrik Schawe}
    \email{hendrik.schawe@cyu.fr}
    \affiliation{Laboratoire de Physique Th\'{e}orique et Mod\'{e}lisation, UMR-8089 CNRS, CY Cergy Paris Universit\'{e}, France}
    \affiliation{Institut f\"ur Physik, Universit\"at Oldenburg, 26111 Oldenburg, Germany}
    \author{Alexander K. Hartmann}
    \email{a.hartmann@uol.de}
    \affiliation{Institut f\"ur Physik, Universit\"at Oldenburg, 26111 Oldenburg, Germany}
    \date{\today}

    \begin{abstract}
        We study the stochastic block model which is often used to
        model community structures and study community-detection
        algorithms. We consider the case of two blocks in regard to its
        largest connected component and largest biconnected component,
        respectively. We are especially
        interested in the distributions of their sizes
        including the tails down to
        probabilities smaller than $10^{-800}$. For this purpose we
        use sophisticated Markov chain Monte Carlo simulations to sample
        graphs from the stochastic block model ensemble.
        We use this data to study
        the large-deviation rate function and conjecture that
        the large-deviation principle
        holds. Further we compare the distribution to the well known
        Erd\H{o}s-R\'{e}nyi ensemble, where we notice subtle differences
        at and above the percolation threshold.
    \end{abstract}

    \maketitle

    \section{Introduction}
    The stochastic block model (SBM) \cite{holland1983stochastic} is a
    generative model for networks with community structure.
    For this purpose, each node is assigned to one of $B$ \emph{blocks}.
    Similar to the Erd\H{o}s-R\'enyi (ER) model \cite{erdoes1960},
    edges between pairs of
    nodes appear with some probability. For the SBM, these probabilities
    can depend on the blocks each node belongs to.
    Thus, the probabilities for edges between or within
    the blocks can be encoded
    in the $B \times B$ \emph{block matrix}. On the one hand this makes the
    model very versatile with an arbitrary number of blocks and arbitrary
    probabilities between the blocks, on the other hand it still stays simple in
    the sense that it is an ensemble of random graphs without any further
    correlations between the edges like the Erd\H{o}s-R\'{e}nyi graph ensemble
    or configuration model \cite{newman_book2010}.
    Indeed in the case of $B=1$ it simplifies to an
    Erd\H{o}s-R\'{e}nyi ensemble.

    In statistical physics there is a persistent interest in the stochastic
    block model as a tool for community detection, i.e., given a network,
    what is the block matrix and to which blocks do the nodes belong most
    probably
    if this realization was drawn from an ensemble of stochastic block models.
    This problem shows interesting behavior as it exhibits two phases: One
    in which a reconstruction of the parameters is possible -- studying
    different approaches how to do that is another active field of studies
    \cite{decelle2011asymptotic,karrer2011stochastic,peixoto2012entropy,Krzakala2013spectral,peixoto2014efficient,peixoto2017nonparametric}
    -- and another phase, where the reconstruction is
    infeasible \cite{decelle2011inference,Nadakuditi2012,darst2014algorithm}.
    In general, the determination of
    community structures is
    algorithmically challenging. This motivated our study, because we
    are interested in whether the detectability is related to the
    simpler properties of the system, like the size of the largest cluster.
    Here, as anticipation of our results, we find
    that we can indeed recognize whether there is some kind of block structure
    present for a small parameter range, especially when also considering the
    far tails of its distribution. But this distinguishability seems to be
    related to the percolation threshold instead of the detectability
    threshold.

    Usually, systems modeled by networks have some kind of functionality, e.g.,
    communication networks enable information exchange between nodes, power
    grids enable power transmission between producers and consumers, and social networks
    exchange, for example, opinions over the edges.
    As a very simple but general indicator of the functionality for sparse
    networks the size $S$ of the largest connected component is useful and
    the most simple global network property of any ensemble. Hence,
    we study here the distribution of $S$ for the SBM. The average
    behavior of $S$ determines the percolation transition. As we will show
    below, the percolation transition of the SBM is related
    to the ER simply by using an effective (average) connectivity.
    This could indicate that the distribution of $S$ is the same for SBM
    and the ER with this effective connectivity. But this is not
    the case, as we will show below, in the percolating region.

    Furthermore, since networks consist of many nodes, which often symbolize entities that
    can fail or vanish, the robustness against this kind of events is of
    relevance. A common idea \cite{albert2000error,Norrenbrock2016fragmentation,Callaway2000network,Cohen2000Resilience,dewenter2015large}
    to measure robustness
    is to remove one or several nodes, either randomly or according to
    ``attack'' rules, and measure its impact on the functionality.
    Here, since we are measuring the functionality in terms of the size of the
    largest connected component, we also
    measure the robustness in terms of the
    size of the largest biconnected component, i.e., the subgraph that will
    stay connected if any node was removed. Note that this observable is
    not an uncommon choice to determine robustness \cite{Newman2008Bicomponents}.

    We scrutinize these properties in very high detail, i.e., we
    do not only look at their mean size, but we obtain their probability
    distributions over practically the whole support, especially including very
    rare events with a probability of less than $p = 10^{-800}$.
    This is, to our knowledge, the first time the tails of the probability
    density function of these properties are studied for any stochastic block model.
    In large-deviation theory \cite{touchette2009}, many probability
    distributions have a special
    shape which allows to remove the leading finite-size influence and describe
    the distributions by the so-called \emph{rate function}.
    As we will show below, here we find a
    comparatively fast convergence of the empirical rate functions calculated
    from the finite-size distributions. This enables us to observe the complete
    large deviation rate function almost directly and conjecture that the
    \emph{large-deviation principle} \cite{touchette2009} holds for this distribution.

    The motivation to study these properties in such great detail is mainly
    fundamental interest in the behavior of these ensembles. The deep
    tails, which we explore here too, should lead to a deeper understanding
    of fringe cases.
    Also, we hope that our numerical high-precision studies
    motivates analytical work in this direction.

    The latter seems possible as the behavior of
    the studied observables
    is known analytically for the related ER ensemble
    \cite{biskup2007large}. Corresponding large-deviation results
    were also obtained using simulational techniques
    \cite{Hartmann2011large,Schawe2019large}.
    Since the ER is a special case of the stochastic block model and is in general
    a good \emph{null-model} to compare other graph ensembles to, we compare
    and contrast it to the SBM. We even show results for the ER ensemble
    for larger sizes than studied before in Refs.~\cite{Hartmann2011large,Schawe2019large}.

    As mentioned in the last paragraph, we previously studied similar
    subjects. Starting with Ref.~\cite{Hartmann2011large}, motivated by an analytical
    expression for the rate function for the size of the largest connected
    component for ER \cite{biskup2007large} one of us compared this expression
    with measurements obtained from simulations and found a rather fast convergence
    of the measurements to the rate function valid in the asymptotic limit, as
    even estimates obtained from graphs with a few hundred nodes showed already
    a very good convergence. Also in the same publication a similar analysis
    is performed for regular lattices instead of ER. Later, in Ref.~\cite{hartmann2018distribution}
    the distribution of the diameter for ER was obtained similarly and
    in Refs.~\cite{Hartmann2017Large,Schawe2019large} we extended
    these results to the distribution of the size of the $2$-core and the
    biconnected component. In the latter we also studied another
    graph ensemble, the famous Barab\'{a}si-Albert ensemble of scale--free
    graphs \cite{barabasi1999emergence}. Here, we extend this line of work
    further by examining the connected component and biconnected component
    on yet another ensemble of random graphs: a simple case of the SBM.

    This case study on a few parameter values of a simple SBM can surely not
    be generalized to all SBMs. So the results we will show in this manuscript
    are primarily applicable to the parameters studied. However, often the
    results give insight into the mechanism leading to a specific behavior,
    in which case we will make educated guesses for which more general cases
    we expect to observe similar phenomena.

    \section{Models and Methods}
    A \emph{graph} is a tuple $G = (V, E)$ of a set of \emph{nodes} $V$
    and a set of \emph{edges} $E$. The number of nodes $\abs{V} = N$ is
    called the \emph{size} of the graph.
    Here we will only scrutinize \emph{undirected}, \emph{simple}
    graphs, i.e., $E \subset V^{(2)} \setminus \{\{u,u\} | u \in V\}$.
    For each node the number of incident edges is its \emph{degree}.
    Since graphs are used to model relations between objects, one of the most
    fundamental properties of graphs is their connectedness.
    Fundamentally, only nodes $i, j$ which are \emph{connected}
    via a \emph{path}, i.e., a sequence of edges $\{\{i, u_1\}, \{u_1, u_2\}, \ldots, \{u_m, j\}\}$,
    can interact at all with each other. The maximal subsets whose members are
    connected are called \emph{connected components}, their \emph{size} is
    the number of elements. It is therefore of interest if a given graph is
    connected, or what the size of its largest connected component is.

    The functionality of a network is for many applications directly dependent
    on a large connected component. For example in a power delivery network
    -- in the best case -- every producer could pass its power to any consumer,
    in a communication network it is desirable that every member can communicate
    with any other member, in a network encoding physical contacts between
    subjects, small connected components would be advantageous to inhibit
    the spreading of disease. While in all these cases maybe other observables
    might capture the functionality better, the size of the largest connected
    component is a reasonable first approximation. In the following we will
    mainly consider its relative size $S$.

    As a second observable we take a look at the closely related
    \emph{biconnected components}, which are the maximal subsets whose members
    are connected by two node-independent paths. This means that one can remove
    any node from a biconnected component and the remainder will still be a
    connected component. The relative size $S_2$ of the largest biconnected component is
    therefore the most simple quantity to judge the \emph{robustness} against node
    removal or failure of a network.

    Algorithmically, one can determine the size of all connected and
    biconnected components in time $\mathcal{O}(\abs{V} + \abs{E})$ by
    performing one modified depth first search on a given graph \cite{Hopcroft1973algorithm,cormen2009introduction,lemon}.
    Note that a node can be part of two distinct biconnected components, such
    that the sum of the sizes of all biconnected components might be larger
    than $N$.

    \subsection{Graph ensembles}
    The Erd\H{o}s-R\'{e}nyi graph (ER) is probably the simplest and first studied
    random graph ensemble \cite{erdoes1960}. It consists of $N$ nodes and any
    possible edge exists independently from all other edges with a probability
    of $p$. If one is interested in sparse graphs, it is convenient to
    parametrize the ensemble with the \emph{connectivity} $c = Np$, which is equal to
    the expected degree.
    In particular, the ER ensemble shows a phase transition from a forest-like structure
    with connected components of size $\mathcal{O}(\log N)$ to a structure with one
    giant connected component of size $\mathcal{O}(N)$ when increasing $c$
    above the critical threshold of $c_c = 1$ \cite{erdoes1960}. Note that beyond the same
    threshold $c_c = 1$ a giant biconnected component of size $\mathcal{O}(N)$ arises \cite{Newman2008Bicomponents}.

    The stochastic block model (SBM) is a random graph ensemble in which every
    node belongs with probability $P_b$ to \emph{block} $b$. Similar to the ER the edges
    exist independently with a fixed probability, but in the SBM the probability
    of the edge $\{i, j\}$ to exist, depends on the blocks $a, b$ of which $i$
    and $j$ are members of, i.e., $p_{ab}$. The diagonal of this block matrix
    governs how tightly connected the nodes within a block are, and the
    off-diagonal elements determine how tightly the connections between
    distinct blocks are, e.g., if the diagonal is zero, every realization will
    be bipartite. If the diagonal elements are larger than the off-diagonal,
    the SBM is called \emph{assortative}; if the off-diagonal elements
    are larger than the diagonal, it is called \emph{disassortative}.
    Note that a homogeneous $p_{ab} = p$ is equivalent to
    the ER. Since we will study sparse SBM, we will parametrize the ensemble
    with connectivities
    \begin{equation}
        c_{ab} = N p_{ab}. \label{eq:edge_prob}
    \end{equation}

\begin{figure}[htb]
        \centering
        \subfigure[\label{fig:sbm:10}]{
            \includegraphics[scale=0.10]{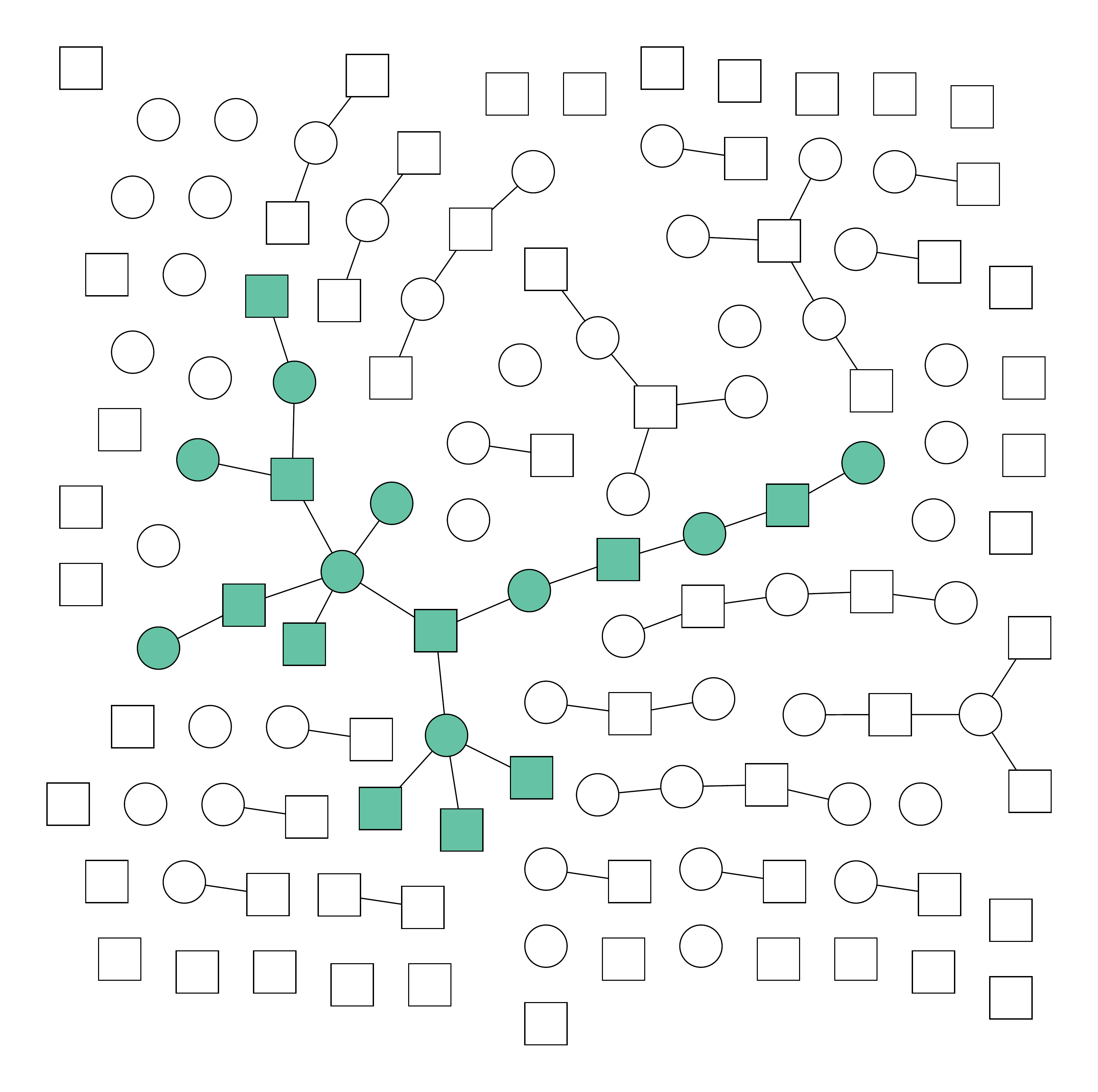}
        }
        \subfigure[\label{fig:sbm:-10}]{
            \includegraphics[scale=0.10]{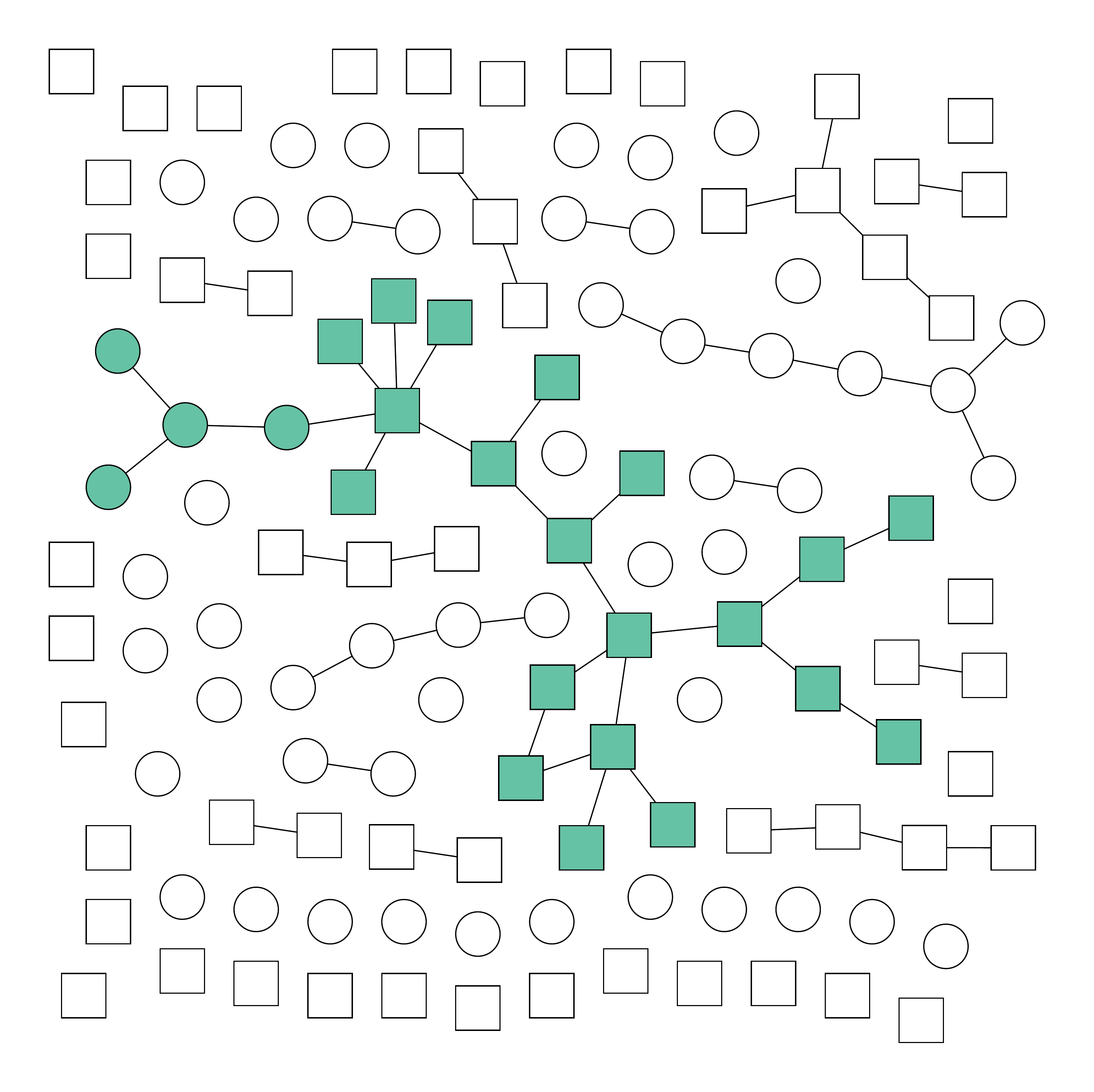}
        }
        \subfigure[\label{fig:perc}]{
            \includegraphics[scale=1]{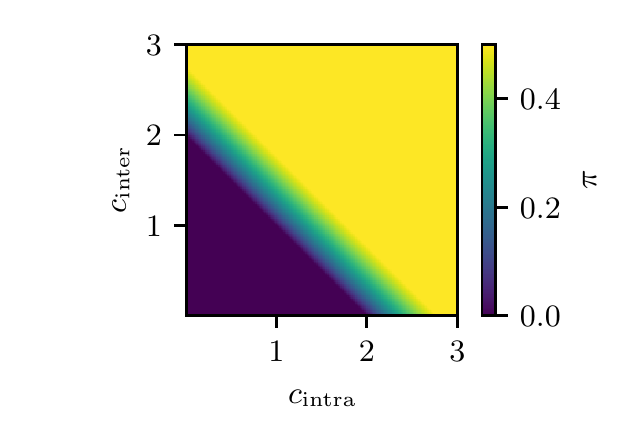}
        }

        \caption{\label{fig:sbm}
            Two example realizations of the SBM with size $N=128$ with two
            blocks (shape of nodes) of equal probability $P_a = P_b = 0.5$.
            The panels show realizations with different connectivities at the
            percolation threshold
            \subref{fig:sbm:10} $c_\mathrm{intra} = 0.1, c_\mathrm{inter} = 1.9$
            \subref{fig:sbm:-10} $c_\mathrm{intra} = 1.9, c_\mathrm{inter} = 0.1$.
            The largest connected components are visualized with colored symbols.
            \subref{fig:perc} numerical solution of Eq.~(\eqref{eq:perc})
            showing the percolation threshold of our particular SBM model.
            Note that the lightest shade does also signify sizes $\pi > 0.5$.
        }
    \end{figure}

    We want to perform a very in-depth study of an SBM ensemble, therefore we will
    treat the most simple special case of SBM, the \emph{planted partition}, i.e.,
    all blocks have the same intra-block (diagonal) connectivity $c_\mathrm{intra}$ and the same
    inter-block (off-diagonal) connectivity $c_\mathrm{inter}$. Further we mainly handle the simplest case
    which is distinct from ER, i.e., $B=2$ blocks, but later on we also show some results for $B=3$. Figure~\ref{fig:sbm} shows
    two examples for different values of $c_\mathrm{inter}$ and $c_\mathrm{intra}$.

    The phase transition where a giant connected component of size $\mathcal{O}(N)$ arises
    happens for $B=2$ at $\brac{c_\mathrm{intra} + c_\mathrm{inter}}/2 = 1$ (cf.~Eq.~\ref{eq:perc_threshold}).
    One can derive this threshold by estimating the size of the largest connected component
    analogously to a method for ER~\cite{hartmann2005phase}.

    We estimate the
    size of the largest connected component by considering the probability $\pi$
    that a randomly selected end-node of a randomly selected
    edge is connected via other edges with the giant component of the graph.
    This means that $\pi$ is the relative size of the largest connected component
    in the asymptotic limit. Note that for the simple ER model with connectivity $c$,
    the probability $q_d$ for degree $d$ of a random-end node of
    a random edge is given by $q_d=dp_d/c$, where $p_d$ is the Poisson
    degree distribution.
    Here, apart from the block memberships, there is no correlation in
    an SBM. We consider the case of
    two blocks $a$ and $b$ with the same probability
    $P_a = P_b = 1/2$ and a symmetrical block matrix.
    To derive the probability $p_d$, we first consider the probability $\tilde p_{d_1,d_2}$
    that a node has degree $d_1$ for connections within the same block and degree $d_2$
    for connections to nodes of the other block. Since both distributions are Poissonian,
    with Eq.~\eqref{eq:edge_prob} and because the size of each sub graph is
    just $N/2$, such that the expected number of neighbors in each sub graph
    is $c_\mathrm{intra}/2$ and $c_\mathrm{inter}/2$, respectively, we obtain
    \begin{equation}
      \tilde p_{d_1,d_2} = e^{-c_\mathrm{intra}/2} \frac{(c_\mathrm{intra}/2)^{d_1}}{d_1!}
      e^{-c_\mathrm{inter}/2} \frac{(c_\mathrm{inter}/2)^{d_2}}{d_2!}\,.
    \end{equation}
    The probability of the total degree $d$, we obtain by summing over all possible combinations which
    sum up to $d$:
    \begin{align}
        p_d = &\sum_{k=0}^d \tilde p_{k,d-k} \nonumber \\
            = & e^{-c_\mathrm{intra}/2-c_\mathrm{inter}/2} \sum_{k=0}^d
                \frac{(c_\mathrm{intra}/2)^{k}}{k!}
                \frac{(c_\mathrm{inter}/2)^{d-k}}{(d-k)!}  \nonumber \\
            = & e^{-c_\mathrm{intra}/2-c_\mathrm{inter}/2} \frac{1}{d!} \sum_{k=0}^d { d \choose k }
                (c_\mathrm{intra}/2)^{k}
                (c_\mathrm{inter}/2)^{d-k} \nonumber \\
        \label{eq:degree}
            = & e^{-c_\mathrm{intra}/2-c_\mathrm{inter}/2} \frac{(c_\mathrm{intra}/2+c_\mathrm{inter}/2)^d}{d!}\,,
    \end{align}
    where we have used the Bionomial formula $(a+b)^d = \sum_{k=0}^d {d \choose k} a^k b^{d-k}$.
    Thus we obtain, quite intuitively,
    the original Poissonian distribution with effective connectivity
    $c=c_\mathrm{intra}/2+c_\mathrm{inter}/2$ being the average connectivity.
    Hence, for the case of $B$ blocks of equal size, one would
    therefore still see the standard Poissonian distribution with effective
    $c=(c_\mathrm{intra}+(B-1) c_\mathrm{inter})/B$. More general cases are straightforward to obtain.

    Therefore, to obtain the percolation threshold, one can proceed as for the standard case, which we recap
    very briefly for completeness. We look at
    $1-\pi$, the probability that a node reached by an edge
    is not connected to the giant component.
    Under the assumption that small components are tree-like, which is true in this
    case for the same reason as for the well-known ER case, $1-\pi$
    can be determined self consistently: Since the probability for each
    neighbor to be not connected to the giant component via its other edges is again $1-\pi$.
    This means $1-\pi = q_1 + (1-\pi) q_2 + (1-\pi)^2 q_3 + \ldots$.
    Using $q_d=dp_d/c$ and inserting Eq.~\eqref{eq:degree} into this expression we get
    \begin{align}
        \nonumber
        1 - \pi =& \sum_{d=1}^\infty e^{-c} \frac{c^{d-1}}{(d-1)!} (1-\pi)^{d-1}\\
        \label{eq:perc}
                =& e^{-\pi c} \,.
    \end{align}
    From this expression we can derive the percolation threshold, since solutions $\pi > 0$ (besides
    the trivial solution $\pi = 0$) become possible for $c > 1$, such that the
    percolation threshold is
    \begin{align}
        \label{eq:perc_threshold}
        B = c_\mathrm{intra} + (B-1) c_\mathrm{inter}.
    \end{align}

    Also, Eq.~\eqref{eq:perc} can be easily solved numerically to estimate the relative size $\pi$ for arbitrary
    connectivity parameters. A heatmap of the solutions for $\pi$ is shown in Fig.~\ref{fig:perc}
    where the transition from a size of $\pi = 0$ to $\pi > 0$ is clearly visible, note the
    symmetry due to the symmetric dependence on just $c=c_\mathrm{intra}/2+c_\mathrm{inter}/2$.
    For intuition, consider the following three edge cases:
    If $c_\mathrm{intra} = c_\mathrm{inter} > 1$, this reduces to
    the well known ER case.
    If $c_\mathrm{inter} = 0$ and $c_\mathrm{intra} > 2$, each block behaves
    like an independent ER graph with $c>1$, such that inside each
    block giant components of size $O(N)$ form.
    If $c_\mathrm{inter} > 2$ and $c_\mathrm{intra} = 0$ a bipartite giant
    component of size $O(N)$ arises.

    \subsection{Large deviations and sampling method}
    We are interested in the whole probability distributions of the
    above mentioned observables. This includes
    additionally to the common events, which are often well characterized by
    the mean and variance, also the tails of the distribution characterizing
    extremely rare events. An especially important class of distributions,
    which is said to obey the \emph{large-deviation principle}, consists of
    distributions parametrized by $N$, here the size of the graph, with a
    probability density function $P_N(S)$ which can be expressed in terms of
    a \emph{rate function} $\Phi(S)$,
    such that $P_N(S) = \exp\left(-N \Phi(S) + o(N)\right)$
    \cite{touchette2009}. Thus, $\Phi(S)$ is independent of $N$ and the
    leading term in $N$ is characterized by the rate function. If such a rate
    function $\Phi$ exists, it means that the tails of the distribution decay
    exponentially in $N$ and $\Phi$ governs how fast exactly
    the tails of the distribution decay. If it has a single
    minimum and is twice differentiable, typical events can be approximated as
    Gaussian distributed for large $N$ \cite{bryc1993remark}.
    Clearly, this principle is not valid for every distribution, since they
    could decay slower or faster than exponential, or exhibit singular behavior.

    Therefore, we want to study whether a large deviation principle holds for the
    distribution of the size of the largest connected component for
    this simple case of SBM. Since the tool of our study is the computer
    simulation \cite{practical_guide2015}, we can only treat realizations of finite size $N$,
    such that we can only obtain the \emph{empirical rate function}
    $\Phi_N(S) = -\frac{1}{N}\ln(P_N(S))$ for multiple sizes $N$. If we observe
    that the empirical rate functions for different sizes converge to a
    limit shape, we assume that this limit shape is the actual rate function
    and that the large-deviation principle is valid here.

    The main idea to obtain the empirical rate functions,
    which include information for extremely rare events, is to perform
    a suitably tailored
    Markov chain Monte Carlo simulation in the space of random graphs.
    Thus, the graphs are not sampled independently but it allows one
    to obtain data of the extremely rare and atypical events.
    In the next chapter we will see, that the distributions of the size of the
    largest connected component of the SBM often have a pronounced multi-peak
    structure. This led us to use Wang-Landau's method (WL) \cite{Wang2001Efficient,Wang2001Determining},
    which is especially suited to overcome valleys in the distribution (or energy landscape).
    Such valleys turned out to be problematic for other methods employed
    previously by the authors \cite{Hartmann2011large,Hartmann2017Large,Schawe2019large}.
    These valleys in the distribution were the main hindrance for larger system sizes
    in previous studies on the ER ensemble \cite{Hartmann2011large,Schawe2019large}.
    So, using WL sampling, we could improve considerably on the size of the
    studied ER graphs. Also, since the multi-peak structure is even more pronounced
    for the SBM, WL is the enabling factor for this study.

    To sketch the idea of WL, consider first that an estimate $g(S)$ of the actual distribution,
    which we are searching for, was known in the beginning of the simulation.
    Then one could construct a Markov chain of random graphs $G$
    using the Metropolis-Hastings algorithm
    with an acceptance probability to change from graph $G$ to $G'$ of
    $p_\mathrm{acc}(G \to G') = \min\left\{ 1, \frac{g(S)}{g(S')} \right\}$
    depending on the observables $S=S(G)$ and $S'=S(G')$ of interest.
    If the estimate is very close to the actual distribution,
    a histogram $H(S)$ of the values encountered during this Markov chain
    would be very flat, i.e., all bins would have about the same number of
    entries. We can then use the deviations from flatness to improve our
    estimate $P(S) \approx g(S) H(S)/\avg{H}$ \cite{Dickman2011Complete}, where
    $\avg{H}$ is the mean count of all bins. This procedure is called
    \emph{entropic sampling} \cite{Lee1993Entropic}, fulfills detailed balance and will
    therefore converge to the correct searched for distribution. The drawback
    is that it may converge very slowly depending on the quality of the
    initial guess $g(S)$.

    The ingenious idea of WL is to get an estimate for $g(S)$ by using the flatness
    of an auxiliary histogram as a criterion to change $g(S)$ during the evolution
    of the Markov chain. Therefore every time an energy $S^*$ is visited, the estimate is
    updated $g(S^*) \mapsto f \cdot g(S^*)$ using the \emph{refinement} factor $f$,
    which is usually initialized as $f=\exp(1)$ and reduced as soon as the
    histogram fulfills some flatness criterion \cite{Wang2001Determining} or
    some set amount of change attempts was performed \cite{Belardinelli2007Fast}.
    First, we need to define the \emph{Monte Carlo time} $t$ in sweeps, i.e.,
    $N$ change proposals.
    Here we use the schedule of Ref.~\cite{Belardinelli2007Fast,Belardinelli2007theoretical},
    where the logarithm of the refinement factor $\ln{f}$ first decreases exponentially
    $\ln{f} \mapsto \ln{f}/2$ every time each bin of the auxiliary histogram
    was visited at least once, this is checked every 1000 $t$. If this criterion
    is fulfilled, the auxiliary histogram is reset. As soon as $\ln{f} < 1/t$, $\ln{f}$
    is decreased as a power law after every sweep $\ln{f} \mapsto 1/t$.
    The algorithm stops as soon as $\ln{f}$ reaches a defined value of
    $\ln{f_\mathrm{final}}$, chosen as $10^{-5}$ in this study.

    Since this means that $p_\mathrm{acc}$ is time dependent, detailed balance
    does not hold and systematic errors might be introduced.
    Therefore we subsequently perform entropic sampling, which is theoretically
    sound, to remove any systematical error.

    This technique depends on the choice of the histogram, such that one has to be
    careful when choosing the binning. Here, fortunately, we have a discrete
    problem, since the size of largest component is the number of nodes, such that
    we can choose a perfect binning of $N$ uniform bins, which does not introduce
    any discretization error.

    One can parallelize WL by performing it independently
    in multiple \emph{windows} and matching the resulting estimates using overlaps
    of the windows. We use up to 14 overlapping windows for this purpose.
    For each window $3\cdot 10^5$ sweeps are simulated: $10^5$ sweeps to reach
    $\ln{f_\mathrm{final}} = 10^{-5}$ and $2 \cdot 10^5$ sweeps of entropic sampling.
    Per window this takes for the largest simulated
    sizes around 40 hours ($N=2048$ for $S$) to 70 hours ($N=1024$ for $S_2$)
    on relatively modern hardware \footnote{Intel Xeon E5-2650 v4},
    fluctuating by approximately $50\%$ in both directions depending on the
    connectivity of the graphs and the acceptance rate.

    One of the most crucial aspects of any Markov chain Monte Carlo simulation
    is the choice of \emph{change move} to generate new trial graphs for
    the chain. Beyond the block membership all edges are independent in the SBM,
    just like the ER. Therefore we create a new trial graph $G'$ by selecting a
    node $i$ in the current graph $G$ at random, removing all of its
    edges and deciding for each other node $j$ randomly whether edge $\{i,j\}$
    is inserted with the appropriate probability depending on their block
    memberships. This change move is ergodic and works reasonably well.

    \section{Results}

\begin{figure}[htb]
        \centering

        \subfigure[\label{fig:ld_dist:p}]{
            \includegraphics[scale=1]{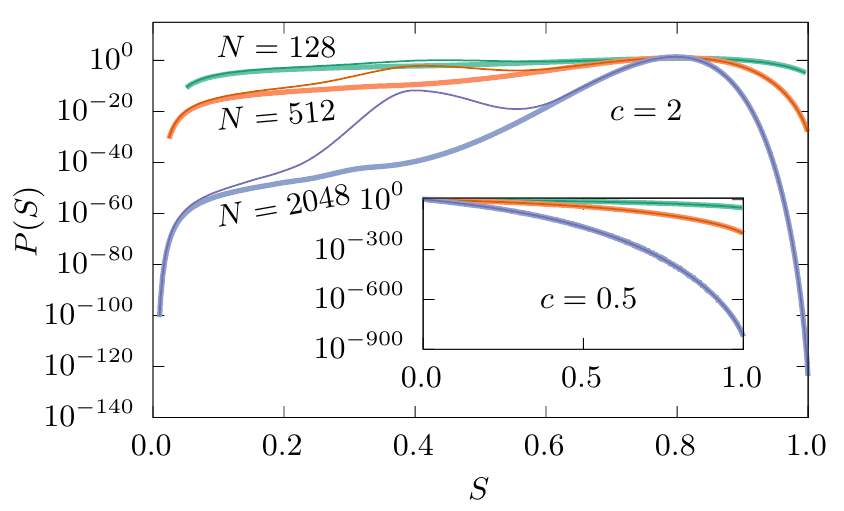}
        }
        \subfigure[\label{fig:rate:multi}]{
            \includegraphics[scale=1]{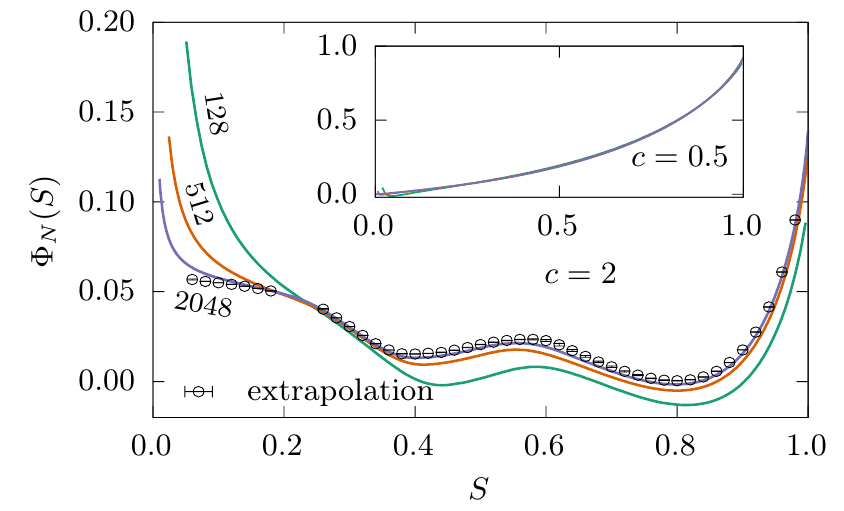}
        }

        \caption{\label{fig:ld_dist}
            \subref{fig:ld_dist:p} Distributions in logarithmic scale
            of the relative size of the
            largest component $S$ for ER (thick lines) and SBM (thin lines)
            for different graph sizes over (almost) the full support.
            \subref{fig:rate:multi} Corresponding empirical rate functions. A
            fast convergence to a limiting shape is observable.
            Main plots show $c=2$, respectively, $c_\mathrm{inter} = 0.1, c_\mathrm{intra} = 3.9$
            and the insets show $c=0.5$, respectively, $c_\mathrm{inter} = 0.1, c_\mathrm{intra} = 0.9$.
            In the main plot data for finite sizes is extrapolated and
            the extrapolation as the estimate of the rate function is
            shown with symbols.
        }
    \end{figure}

    In Fig.~\ref{fig:ld_dist:p} we show for some system sizes
    the resulting distributions
    for the cases of low connectivity $c=0.5$ ($c_\mathrm{inter}=0.1,
    c_\mathrm{intra}=0.9$) in the non-percolating regime (inset)
    and of higher connectivity $c=2$ ($c_\mathrm{inter}=0.1,
    c_\mathrm{intra}=3.9$) in the percolating regime.
    Here, and in the following, data for the SBM is visualized with
    thin, dark lines and for the ER with thick, lighter lines.
    If only one line is visible, the data of both ensembles coincide.
    Different shades mark different sizes $N$. Here we see that the SBM
    exhibits in the $c=2$ case a strongly different behavior than the ER. This
    manifests for the largest system size in structures of the probability
    density function (pdf) below probabilities of $10^{-15}$ and would therefore be
    undetectable with conventional methods.

    To foster intuition about the relation of the rate function with the
    probability density, we show the empirical rate functions $\Phi_N(S)$
    obtained from the probability density functions
    in Fig.~\ref{fig:rate:multi}. For clarity only
    data for the SBM is visualized. Here, we observe that
    the smallest size $N=128$ shows non-monotonous and rather large deviations
    but $\Phi_{512}(S)$ and $\Phi_{2048}(S)$ are already very close over large
    parts of the support. The extrapolation is performed point-wise,
    at equal values of $S$ for multiple system sizes. We assume a power law
    with offset $\Phi_N(S) = \Phi_\infty(S) + a N^b$ for the extrapolation, which
    was already used before for this task \cite{schawe2018avoiding} and fits
    quite well to our data. Since different
    system sizes have a different number of bins, we interpolate $\Phi_N$
    linearly for convenience, which should only introduce negligible error
    due to the dense bins. Note that the minimum of the extrapolated estimate of
    the rate function is at $\Phi_\infty(S_\mathrm{min}) \approx 0$.
    The region around $S \approx 0.2$ could not be extrapolated well due to the
    crossing of different system sizes.
    For $c \le 1$ we do not perform this extrapolation since all system sizes
    already yield almost equal rate function estimates.
    For clarity, we will not show the extrapolation in following figures, since it
    always is very close to our data for the largest system size, i.e., $N=2048$
    seems to be large enough that the empirical rate function $\Phi_N$ is
    sufficiently close to our extrapolation $\Phi_\infty$, which we handle as
    an estimate of the actual asymptotic rate function.

    Note that the nature of numerical studies is that we have to rely on the
    problem being well-behaved. While we have sizes which are large enough to
    show a convergence to some value, it is theoretically possible that this
    value is not the asymptotic limit but only a `plateau' and that the
    behavior changes for even larger system sizes.
    Since we know the exact behavior of the rate function
    for the ER and we previously observed a very
    nice and quick convergence of measurements in finite systems with similar system
    sizes to the exact asymptotic form \cite{Hartmann2011large},
    we argue that the simple SBM under scrutiny should behave
    very similarly well-behaved.

    Therefore, we conjecture that the large deviation principle holds
    for $S$ and $S_2$ of this variant of the SBM and is approximated by the empirical
    rate functions for the largest measured sizes shown in the following figures.
    Since we observe this above and below the percolation threshold,
    and because we do not expect any other dramatic changes in the structure,
    this should hold for all parameters $c_\mathrm{inter}$ and $c_\mathrm{intra}$.

    As a side remark, consider a finite temperature $T$ ensemble, where the
    occurrence of realizations was weighted with a Boltzmann weight $e^{-S/T}$,
    treating the size of their largest connected component $S$ as energy,
    studied, e.g., in Ref.~\cite{Hartmann2011large}. The two-peak structure
    corresponds to two transitions of first order at two distinct temperatures
    $T$. At one transition, two large coexisting components appear, at the other
    transition, one single biggest component emerges, see the discussion below.

    \begin{figure}[h!tb]
        \centering

        \subfigure[\label{fig:rate:c05}]{
            \includegraphics[scale=1]{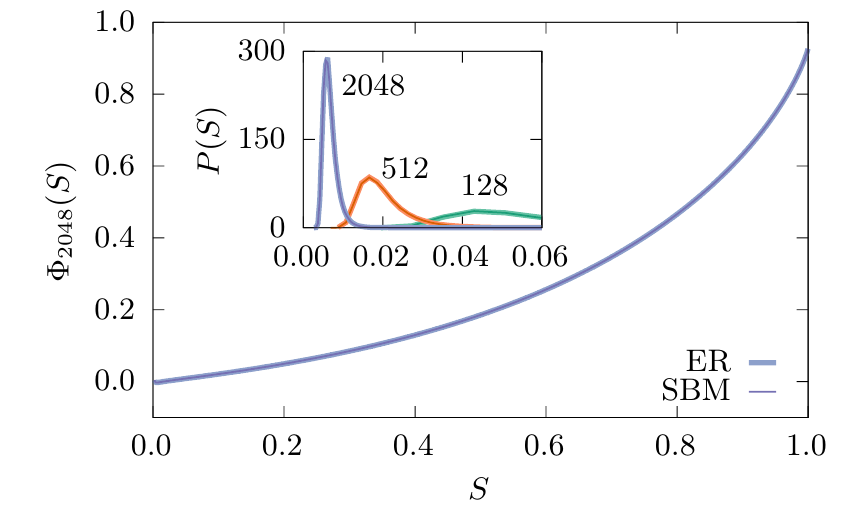}
        }
        \subfigure[\label{fig:rate:c1}]{
            \includegraphics[scale=1]{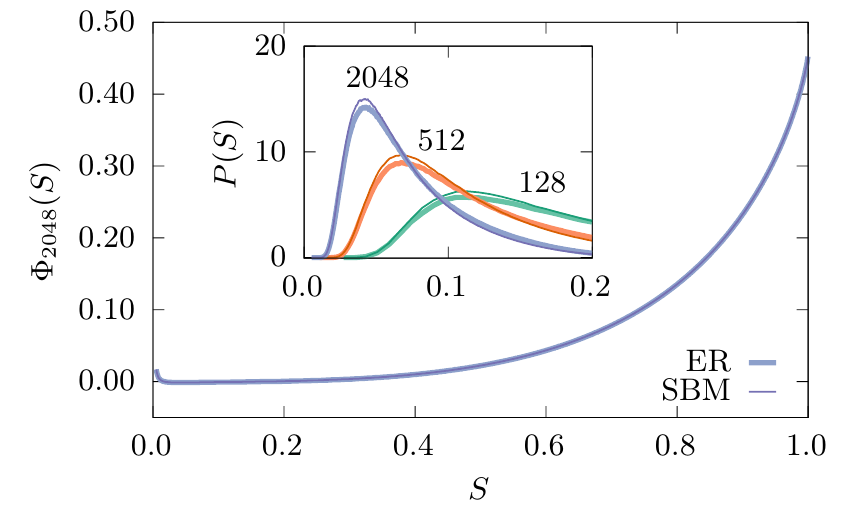}
        }

        \subfigure[\label{fig:rate:c4}]{
            \includegraphics[scale=1]{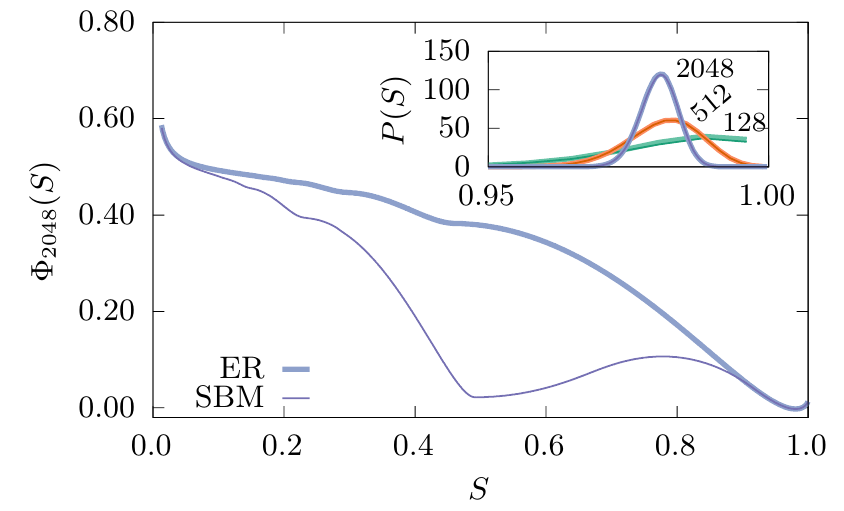}
        }

        \caption{\label{fig:rate}
            The main plots show the empirical rate functions $\Phi_{2048}(S)$ for
            different connectivities of both the ER
            (thick lines) and SBM (thin lines) ensemble, which coincide often:
            \subref{fig:rate:c05} $c = 0.5, c_\mathrm{inter} = 0.1, c_\mathrm{intra} = 0.9$,
            \subref{fig:rate:c1} $c = 0.5, c_\mathrm{inter} = 0.1, c_\mathrm{intra} = 3.9$,
            \subref{fig:rate:c4} $c = 0.5, c_\mathrm{inter} = 0.1, c_\mathrm{intra} = 7.9$.
            Their insets show the corresponding probability density functions for different sizes.
            Note that the normalization for $P(S)$ in the insets
            is such that the area under the curve
            is unity (thus for better comparison we show densities
            although the support is discrete).
        }
    \end{figure}

    In Fig.~\ref{fig:rate} the empirical rate functions and distributions for
    finite system sizes $N$ are shown for different parameter sets, especially
    below and above the percolation threshold.

\begin{figure}[htb]
        \centering

        \subfigure[\label{fig:twocomp}]{
            \includegraphics[scale=0.066]{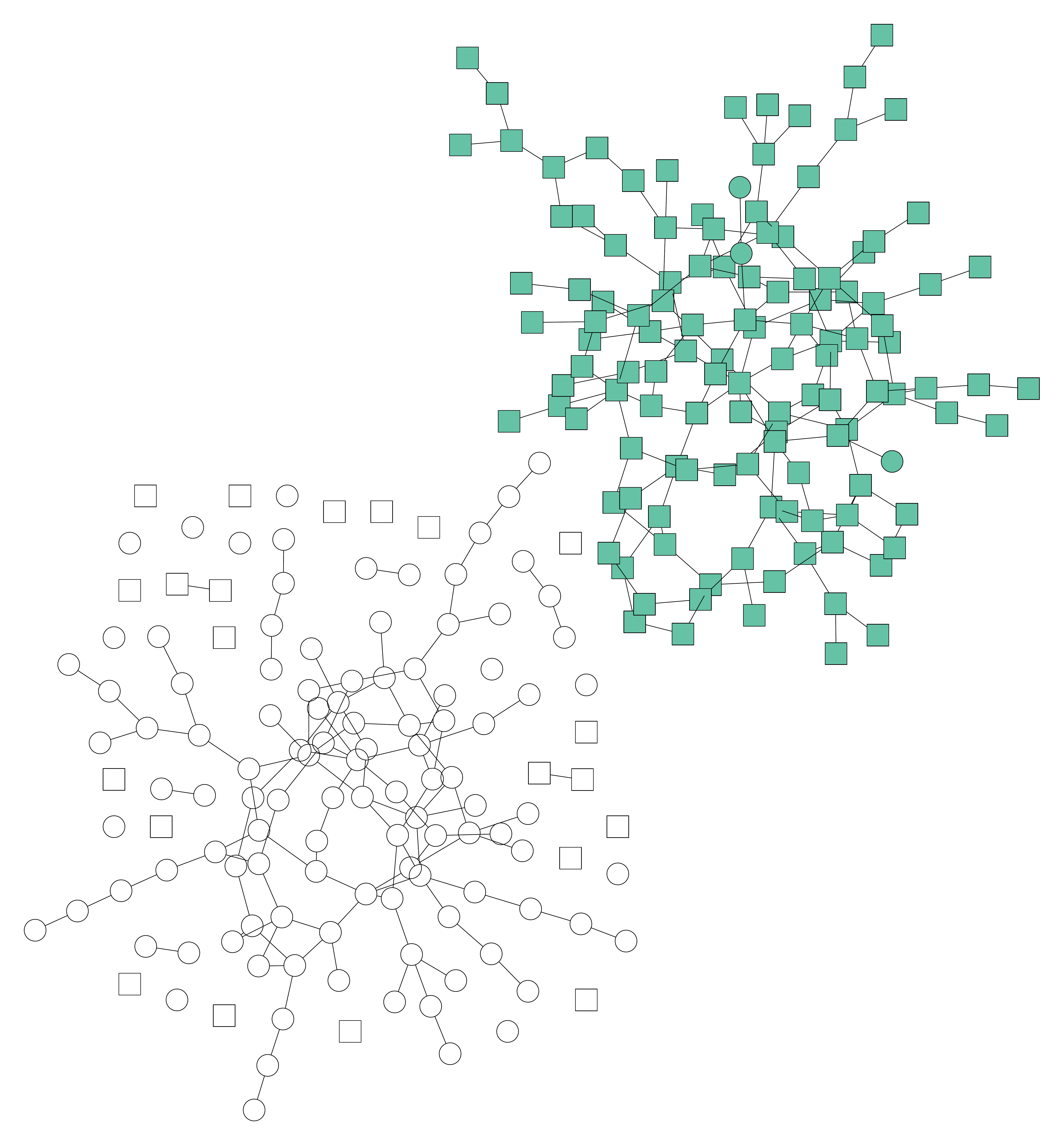}
        }
        \subfigure[\label{fig:onecomp}]{
            \includegraphics[scale=0.066]{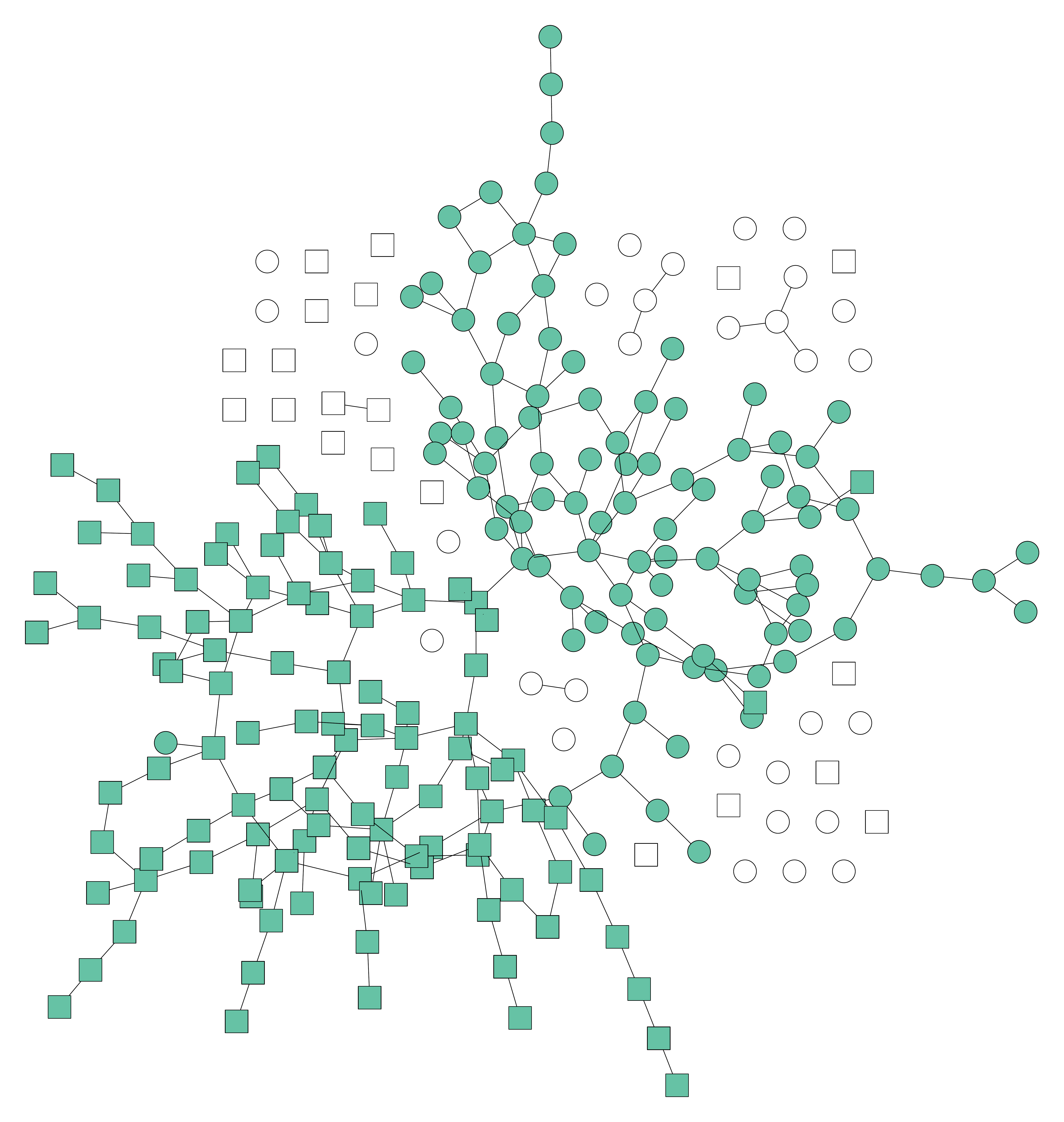}
        }

        \caption{\label{fig:comp}
            Examples for SBM realizations at $c_\mathrm{inter} = 0.1, c_\mathrm{intra} = 3.9, N=256$.
            The two blocks are visualized as nodes of different shapes, the
            largest connected component consists of colored nodes. These are
            realizations originating \subref{fig:twocomp} from the left
            peak and \subref{fig:onecomp} from the right peak.
        }
    \end{figure}

    The peculiar two-peak structure of the rate function of the SBM above the
    percolation threshold in Fig.~\ref{fig:rate:multi} and \ref{fig:rate:c4} can
    be explained rather simple. The left peak, consists of realizations, where two
    separate but large connected components exist -- one in each block.
    For $c = 4$ this peak is at $S \approx 0.5$ since almost all nodes within one
    block are connected. For $c=2$ the connected components within a block are
    smaller, such that we observe this peak at $S \approx 0.4$.
    Figure~\ref{fig:twocomp} shows an example realization of this
    type. Since it is exponentially unlikely that no inter-block edge exists,
    the occurence of this structure is exponentially suppressed, resulting in a value of
    the rate function at this positon larger than zero, and are subsequently
    not visible in the distributions for moderately large systems.
    Due to this mechanism, we expect that
    the left peak will be suppressed at large sizes for all choices of the
    connectivity parameters in the two block planted partition, as long
    as $c_\mathrm{inter} > 0$.
    Also beyond the planted partition, i.e., with different connectivities between
    different blocks, we expect that the second peak is suppressed for large sizes
    as long as all inter-block connectivities are non-zero.
    The main peak at $S \approx 0.8$, respectively $S \approx 1$, on the other
    hand, contains the instances in which
    the connected components inside of the blocks are connected with each other,
    as visualized in Fig.~\ref{fig:onecomp}.

\begin{figure}[htb]
        \centering

        \subfigure[\label{fig:rate:c2}]{
            \includegraphics[scale=1]{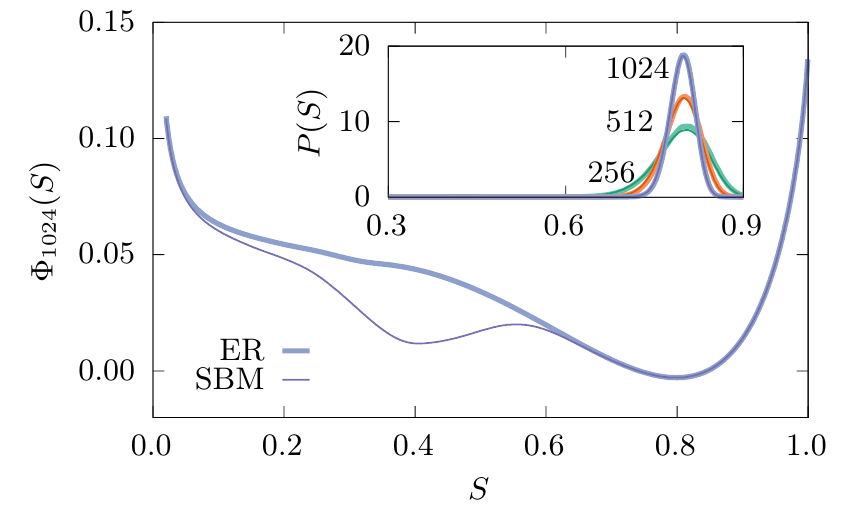}
        }
        \subfigure[\label{fig:rate:bi_c2}]{
            \includegraphics[scale=1]{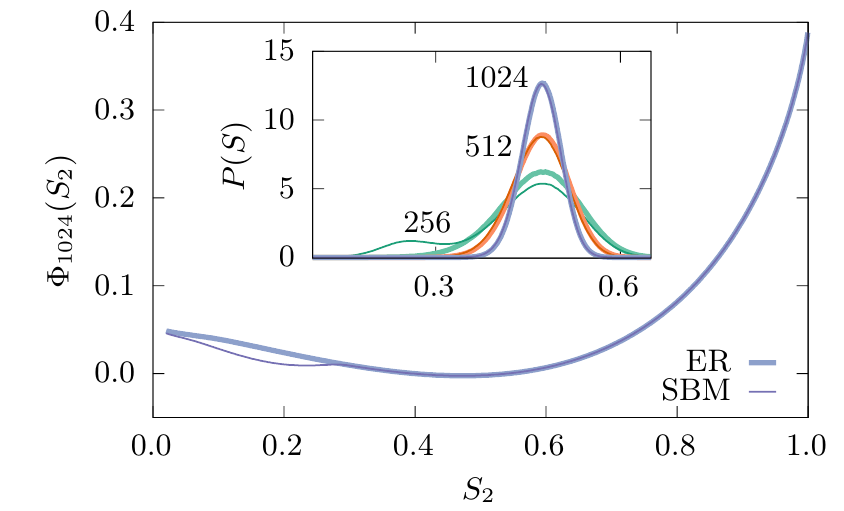}
        }

        \caption{\label{fig:rate2}
            The main plots show the empirical rate functions \subref{fig:rate:c2} $\Phi_{1024}(S)$
            of the relative size of the largest connected component and
            \subref{fig:rate:bi_c2} of the largest biconnected component $\Phi_{1024}(S_2)$
            of both the ER (thick lines) and SBM (thin lines)
            ensemble for $c=2, c_\mathrm{inter} = 0.1, c_\mathrm{intra} = 3.9$. Note that the
            right tail reaches down to $\Phi_{1024} \approx 3.8$, i.e., probabilities of
            $P(S_2) \approx e^{-1024 \cdot 3.8} \approx 10^{-1690}$.
            Their insets show the corresponding probability density for multiple sizes.
            Note that the normalization is such that the area under the curve
            is unity (thus for better comparison we show densities
            although the support is discrete).
        }
    \end{figure}

    Also note that the same two-peak structure exists in the distribution of
    the largest biconnected component visualized in Fig.~\ref{fig:rate:bi_c2}.
    The main mechanism causing the second, smaller peak is the same as for
    the largest connected component: due to the low value of
    $c_\mathrm{inter}$ it is more probable
    that two biconnected components within each of the blocks are not connected by
    two inter-block edges. Since a split in two large biconnected components is
    more probable than the split into two large connected components, we can
    observe the second peak for the smallest size $N=256$ even in the linear scale
    depiction of $P(S_2)$ shown in the inset of Fig.~\ref{fig:rate:bi_c2},
    but not in $P(S)$ in Fig.~\ref{fig:rate:c2}.
    The most probable size of the largest bicomponent is naturally
    smaller than the most probable size of the largest connected component at slightly
    larger than $0.4$, and correspondingly the maximum of the second peak is at half this
    value, slightly larger than $0.2$. The smaller magnitude of the second peak is caused
    at least partially by a much higher probability that there are no large biconnected
    components inside the blocks, which leads generally to a broader distribution.
    Note that the vertical axis spans a far larger range in the diagram for the
    biconnected component, which exaggerates the difference.

    The most striking properties of the distributions $P(S)$ for
    different values of the connectivity is the surprising way they differ
    between ER and SBM. We are able to assess these differences, since
    our large-deviation sampling approach gives us access to the tails:
    Below the percolation threshold in Fig.~\ref{fig:rate:c05}
    the two distributions are visually indistinguishable, in the
    peak (shown in the inset) as well as in the tails (shown in the main plot).
    At the threshold in Fig.~\ref{fig:rate:c1}, one can see significant
    deviations in the peak, but the tails are again indistinguishable.
    Surprisingly, above the threshold in Fig.~\ref{fig:rate:c4} and \ref{fig:rate:c2} the peaks of
    the distributions are again visually indistinguishable, but the tails
    show qualitatively different behavior with a far more pronounced
    second peak for the SBM case.

    For the assortative case $c_\mathrm{inter} \ll c_\mathrm{intra}$, which
    we studied here mainly, it is plausible that the size of the largest connected
    component should differ the most close to the threshold. Since here, close
    to the percolation threshold is the only parameter regime where the inter-block edges do
    matter at all. Far below the threshold, the SBM realization consists of
    trees with members from only one block, but since our observable $S$ does
    not account for the block memberships, this is indistinguishable from ER.
    Far above the threshold the blocks are connected components and
    as long as there are any inter-block edges, the largest connected component
    will typically include almost the whole graph -- the same as the ER case.
    Therefore, in the $c_\mathrm{inter} \ll c_\mathrm{intra}$ case, only around
    the threshold the peaks of the distributions can differ at all.
    Note that these arguments are valid also for higher number of blocks $B$
    and for more general connectivities between the blocks, if they are
    assortative enough.

    In the disassortative case $c_\mathrm{inter} \gg c_\mathrm{intra}$,
    we found that the distribution
    $P(S)$ is generally indistinguishable within our high precision numerical
    data from the ER case, even in the far tails (not shown). This is
    not surprising since the mechanism of two unconnected clusters leading
    to the differences in the assortative cases, can not occur in (almost) bipartite graphs.
    We therefore conclude that the size of the largest connected component
    does differ at most very weakly between this simple SBM and ER.
    For the size of the largest biconnected component the results are
    qualitatively the same and the same arguments apply.

    The balanced case $c_\mathrm{inter} = c_\mathrm{intra}$ is equal to
    the ER ensemble, and therefore trivially does not differ.

    As a more formal method to judge whether or not the peak regions of ER and
    SBM are indistinguishable, we use the \emph{Epps-Singleton} test \cite{epps1986omnibus,scipy},
    which is designed to estimate the probability $p_\mathrm{ES}$ that two
    samples from discrete distributions originate from the same distribution.
    In order to smooth these results, we average the $p$-values obtained for 100
    independent pairs of samples, each containing $10^4$ independent measurements
    of the largest connected components. Note that such a procedure diminishes
    greatly the power of the test, and one would usually choose something like
    Fisher's method to combine independent $p$-values. However, since for
    our purpose, we want to explore a trend for large systems instead
    of analyzing specific cases, the extremely conservative
    $\overline{p}_\mathrm{ES}$ should still be a useful and especially smooth metric.

\begin{figure}[htb]
        \centering

        \subfigure[\label{fig:es}]{
            \includegraphics[scale=1]{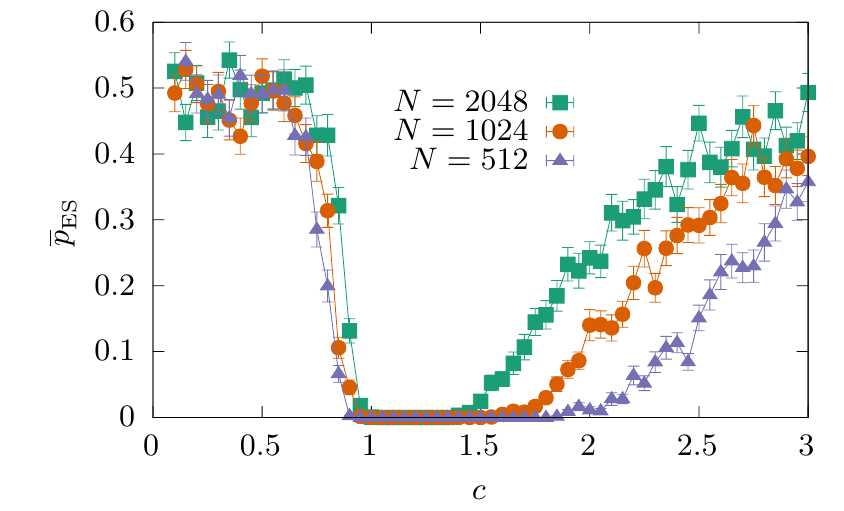}
        }
        \subfigure[\label{fig:heatmap}]{
            \includegraphics[scale=1]{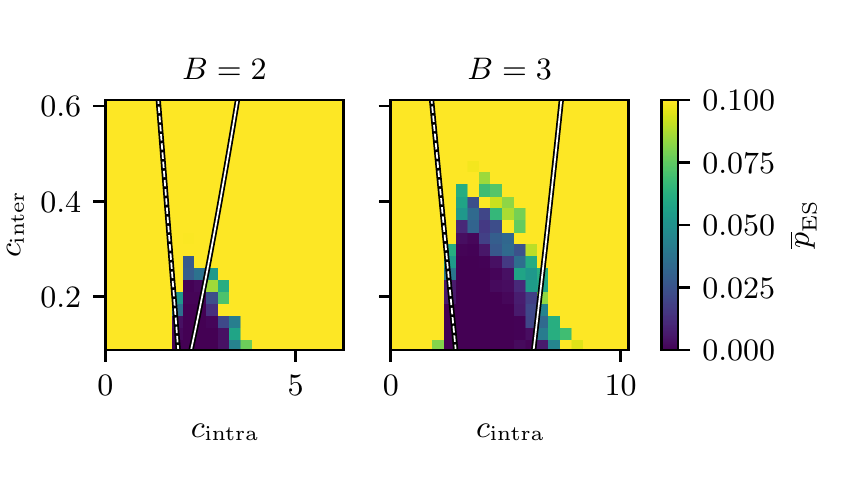}
        }

        \caption{\label{fig:es2}
            \subref{fig:es} Average confidence of $100$ Epps-Singleton tests, that
            two samples of $S$ (with $10^4$ measurements each), one obtained from the ER and
            the other from the SBM with $c_\mathrm{inter} = 0.1, c_\mathrm{intra} = 2c - c_\mathrm{inter}$,
            originate from
            the same distribution. Low values mean that we can surely distinguish
            the two ensembles, high values mean that we can not.
            \subref{fig:heatmap} Heatmaps of the same Epps-Singleton tests at $N = 1024$ for
            more combinations $c_\mathrm{inter}$ and $c_\mathrm{intra}$ for
            both $B=2$ and $B=3$.
            The solid line marks the corresponding community detectability thresholds
            $\abs{c_\mathrm{inter} - c_\mathrm{intra}} = B\sqrt{c}$ \cite{decelle2011inference}
            and the dashed line shows the percolation threshold $B = c_\mathrm{intra} + (B-1) c_\mathrm{inter}$.
            For every pixel $2\cdot10^6$ independent measurements were used, smoothed
            using the method explained in the text. Note that the other branch of the
            threshold for disassortative SBM is not visible due to the short vertical
            axis. However, that region shows solid non-distinguishability.
        }
    \end{figure}

    We used this procedure to estimate $\overline{p}_\mathrm{ES}$
    for multiple values of $c$, in the case of the SBM, we fixed $c_\mathrm{inter}=0.1$
    and varied $c_\mathrm{intra} = 2c-c_\mathrm{inter}$. Figure~\ref{fig:es} shows the result of
    this analysis. Very low values of $\overline{p}_\mathrm{ES}$ signal that
    the two samples originate from different distributions, i.e., the
    distributions are distinguishable. High values, say above the typical $5\%$ mark,
    signal that we can not exclude the possibility that the two samples originate
    from the same distribution. The exact $p$-values
    should not be taken too seriously, since they greatly depend on the number of samples
    and are overestimated due to our smoothing.

    In accordance with our visual
    interpretation above, the distributions for connectivities
    around the transition at $c_c=1$, are distinguishable using this statistical test.
    Especially, the range where the distributions are distinguishable shrinks with increasing
    system size. Also note, that using different statistical tests,
    like Kolmogorov-Smirnov \cite{press2007numerical} or
    Anderson-Darling \cite{anderson1952asymptotic,scipy}, leads to
    similar results (not shown).

    In Figure~\ref{fig:heatmap} we scrutinize for which values of the inter and intra
    block connectivities, the main part of the distribution allows us to distinguish
    ER from SBM, i.e., recognize that there is some community structure in the
    network. There the parameter space which can be distinguished is visualized with
    dark colors. Note that Fig.~\ref{fig:es} suggests that this dark region should
    shrink from the right for larger sizes $N$. With this in mind, this figure
    confirms our observations from before that ER and SBM differ the most around
    the percolation threshold, which is marked by a dashed line. It also shows that
    this distinguishability is only given at low values of $c$. This is expected since
    using the size of the connected component one can not be able to
    distinguish ER from SBM when the connectivities are high enough that the
    giant component almost always contains every single node.

    An interesting question coming to mind is, whether this behavior
    is also related to the transition from detectable community structure to not-detectable
    community structure $\abs{c_\mathrm{inter} - c_\mathrm{intra}} > B\sqrt{c}$ \cite{decelle2011inference}.
    Therefore, we also marked the community detection threshold with a solid line, i.e., realizations right of
    the solid line can be used to reconstruct the block membership of the nodes, e.g., by the sophisticated
    methods of Ref.~\cite{decelle2011inference}.
    For $B=2$ the distinguishability using $P(S)$ and the community detection threshold
    are quite close to the percolation threshold at small values of $c_\mathrm{inter}$.
    The $B=3$ case on the right of Fig.~\ref{fig:heatmap}, especially considering that
    the darkly marked distinguishable region should shrink from the right for larger
    values of $N$ (cf.~Fig.~\ref{fig:es}), shows that the community detection threshold
    is located at considerably higher values of $c_\mathrm{intra}$, than the sparsest
    realizations we can distinguish.
    Thus it appears that we can use $P(S)$ to detect the existence of community structure
    (or some structure absent in ER) in realizations where the actual detection of
    communities is impossible. Interestingly, it seems that realizations, whose
    communities can be reconstructed (right of the solid line), can not be
    distinguished by the typical behavior of $S$. Here systematic, numerically also demanding,
    studies for various larger values of $B$ could possibly be of interest.

    Note that the behavior of the tail, which obviously differs
    for large $c$, e.g., in Fig.~\ref{fig:rate:c2}, is immaterial for this
    statistical test. For analysis of the tail behavior, we will introduce
    the area between the empirical rate functions
    \begin{align}
        A = \int_0^1 \mathrm{d}S \abs{\Phi_N^\mathrm{SBM}(S) - \Phi_N^\mathrm{ER}(S)}
    \end{align}
    as a measure of distinguishability.

\begin{figure}[htb]
        \centering

        \includegraphics[scale=1]{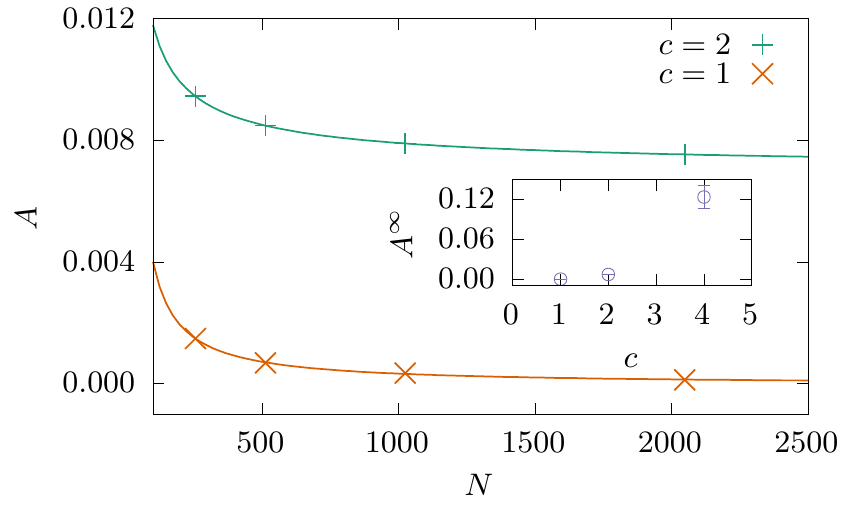}

        \caption{\label{fig:area}
            Area $A$ between empirical rate functions of ER and SBM
            extrapolated using a fit of the form
            $A(N) = A^\infty + aN^b$. The offset $A^\infty$ at a connectivity
            of $c=1$ is compatible with zero at $A^\infty_{1} = 3\cdot 10^{-5} \pm 7\cdot 10^{-5}$,
            i.e., the rate functions of ER and SBM appear to become indistinguishable.
            For a connectivity of $c=2$ we obtain an offset $A^\infty_{2} = 0.007(1)$,
            and for $c=4$ (not shown to preserve detail of the plot) $A^\infty_{2} = 0.12(2)$,
            i.e., the rate functions of ER and SBM appear to stay distinct.
            This behavior $A^\infty(c)$ is shown in the inset.
        }
    \end{figure}

    In Figure~\ref{fig:area} the area $A$ between the empirical rate functions
    is shown for multiple system sizes $N$
    at connectivities of $c=1$ and $c=2$. To estimate whether the differences
    between the empirical rate functions are finite size effects, or persist in the infinite
    limit of the rate function, we extrapolate the area to infinite systems using
    the ansatz $A(N) = A^\infty + aN^b$, which fits quite well to the data.
    We find that for $c=1$ the area $A^\infty$ and therefore the difference vanishes
    within errorbars in the limit of infinite systems. The rate functions for $c=2$,
    on the other hand, stay clearly distinct between ER and SBM. This distinctiveness
    increases for larger values of $c$ (and fixed $c_\mathrm{inter}$), which
    is shown in the inset of Fig.~\ref{fig:area}. This means, that given the far tails
    of the distribution, we can determine the existence of blocks not only at and closely beyond the
    percolation threshold but also far above it.

\begin{figure}[htb]
        \centering

        \subfigure[\label{fig:scatter_sbm}]{
            \includegraphics[scale=1]{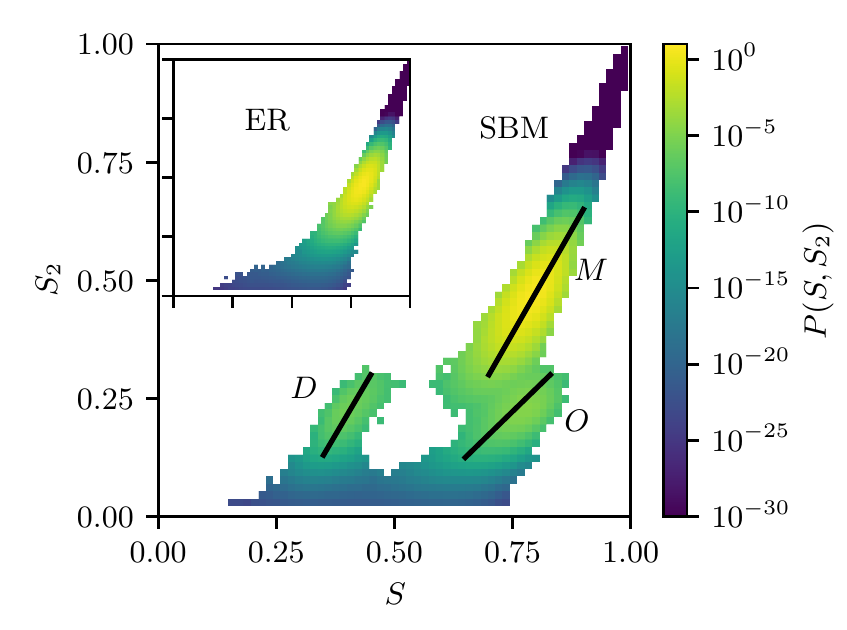}
        }

        \subfigure[\label{fig:ex_a}]{
            \includegraphics[scale=0.063]{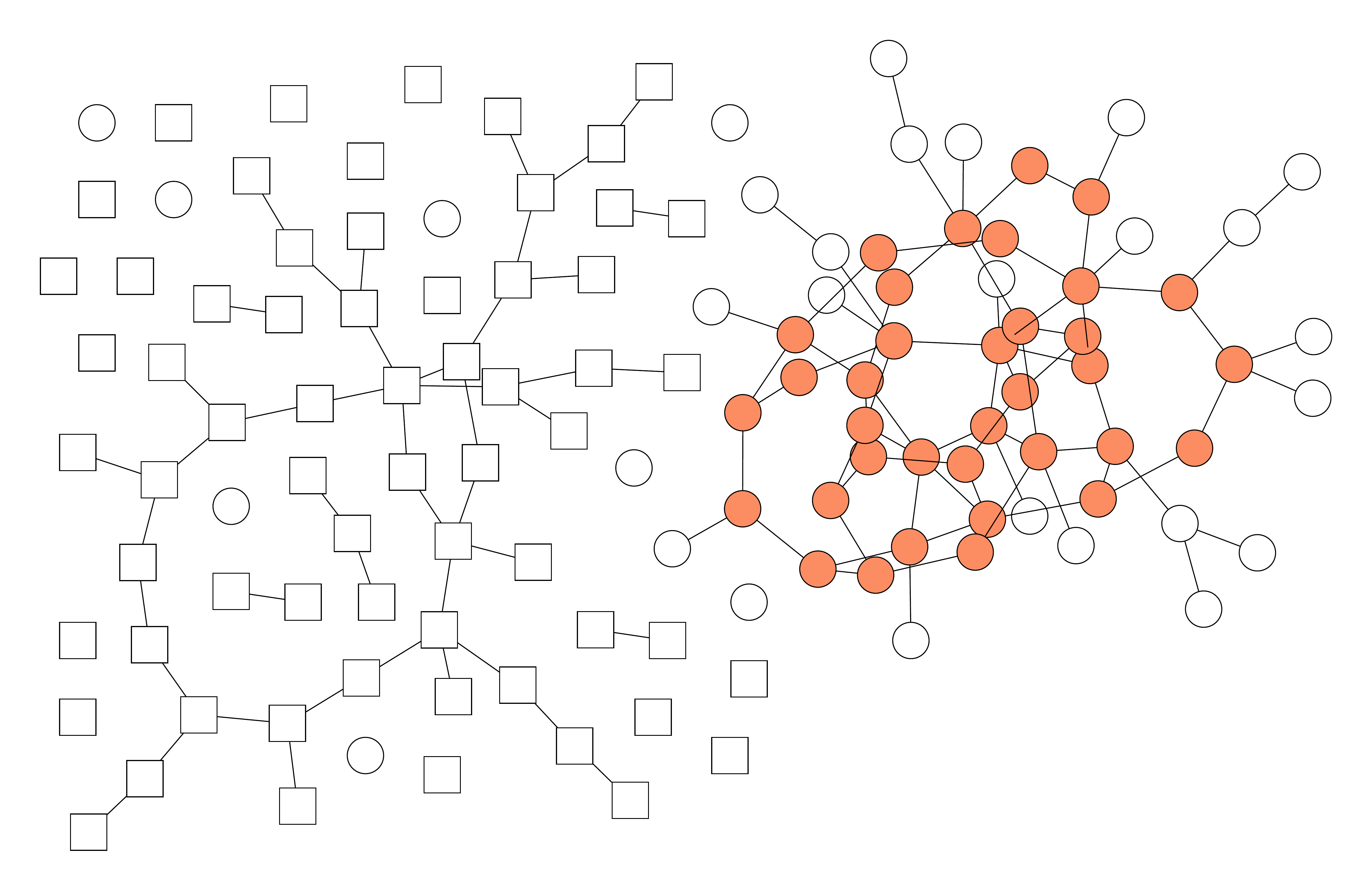}
        }
        \subfigure[\label{fig:ex_b}]{
            \includegraphics[scale=0.063]{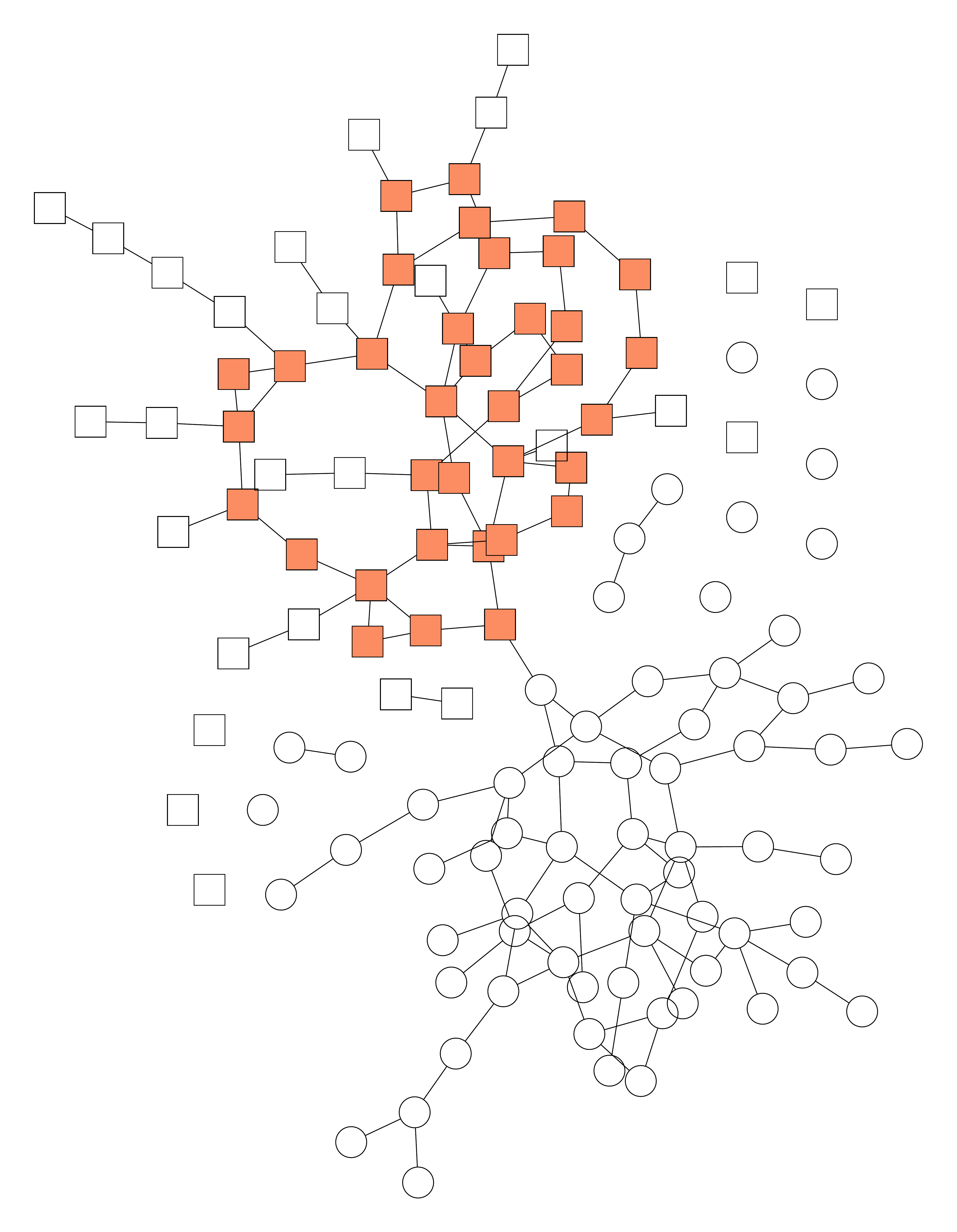}
        }
        \subfigure[\label{fig:ex_c}]{
            \includegraphics[scale=0.063]{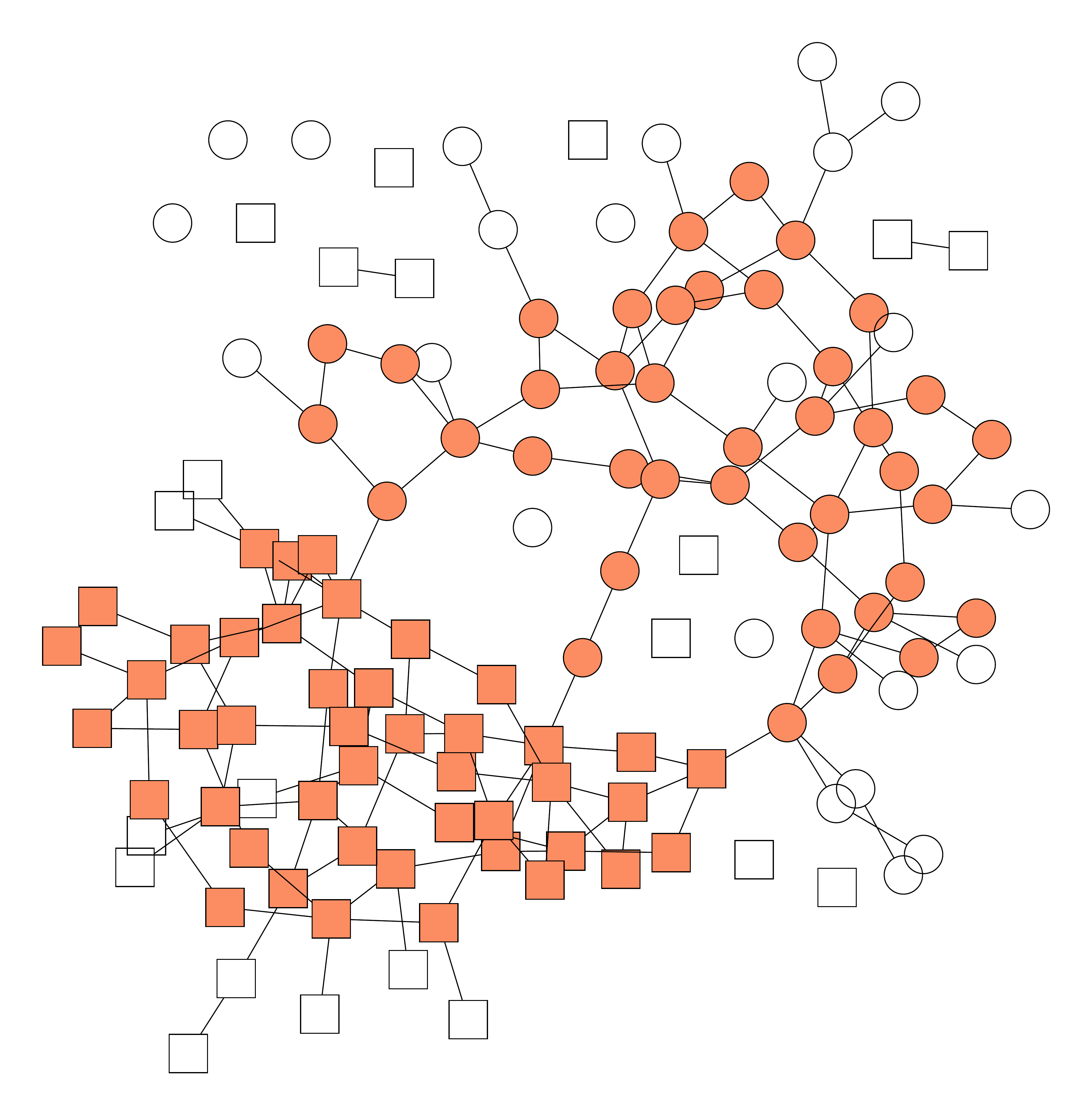}
        }

        \caption{\label{fig:scatter}
            Comparison of parts of the joint probability density $P(S, S_2)$.
            SBM ($c_\mathrm{inter} = 0.1, c_\mathrm{intra} = 3.9$) is shown
            in the main plot, ER ($c=2$) in the inset. The data
            is for $N=1024$ and collected during the entropic sampling of a WL
            simulation of the size of the largest biconnected component.
            The color-scale is compressed to increase the visibility of
            the central structures despite the very large range of probabilities to
            be visualized. For white points no corresponding samples $(S, S_2)$
            were encountered during the entropic sampling. Here, the space is discretized
            into $64 \times 64$ bins.
            The black lines are guides to the eye and indicate the regions
            of highest probabilities. The corresponding labels
            are referenced in the text.
            The three classes identified in \subref{fig:scatter_sbm}
            are illustrated with examples of size $N=128$ with highlighted
            largest biconnected component: \subref{fig:ex_a} example of region D,
            \subref{fig:ex_b} example of region O and \subref{fig:ex_c} example of region M.
        }
    \end{figure}

    To gather insight how configurations with especially large or especially
    small biconnected components look like, we consider the correlations
    between the size of the connected and biconnected components.
    Using Bayes' theorem, we can estimate parts of the joint probability
    density $P(S, S_2)$ from a single WL simulation. Therefore we save
    during the entropic sampling phase
    the pairs $(S, S_2)$ of observables we encountered in the Markov chain and
    estimate the conditioned probability $P(S | S_2)$ from them. This can be
    used with the result of the WL simulation, $P(S_2)$, to obtain a part of the
    joint probability density $P(S,S_2) = P(S | S_2) P(S_2)$.
    Note, that with a much higher numerical effort it would also be possible
    to obtain the full joint probability density using a two-dimensional
    WL variant \cite{Wang2001Determining}.
    In Fig.~\ref{fig:scatter} parts of the joint probability density are shown.
    One notices that the correlations for the SBM above the percolation threshold
    show a surprising multi-modal structure, which is marked and labeled in black.
    However, we will see that this is actually plausible and we will discuss the
    structure of the realizations inside each of the three regions.

    In the region labeled $D$ (divided), which is not
    present in the ER, we see that inside of the highly connected blocks
    of the SBM, which are not yet connected to each other, biconnected components exist (cf.~Fig.~\ref{fig:ex_a}).
    The group of realizations, labeled $O$ (one connection), indicates that there is
    a considerable amount of realizations where already giant connected
    components spanning both blocks exist ($S > 0.5$), but the largest biconnected
    component is still restricted to one of the blocks with $S_2 < 0.3$. These
    are mostly realizations where the connected components inside each block
    are connected by a single edge (or multiple edges arriving at a single node) (cf.~Fig.~\ref{fig:ex_b}).
    Part of this region are also, though less often, configurations with a
    biconnected component inside one block connected to multiple tree like
    structures consisting of nodes of the other block.
    Interestingly both types of configuration coexist in our simulations.
    Since both of these groups rely on the high intra-block connectivity, they
    do not occur in the ER ensemble.

    In the region labeled $M$ (multiple connections) of Fig.~\ref{fig:scatter}, which also occurs for the ER,
    one sees perfect correlation between the size of the two
    types of components. The larger the biconnected component should be, the
    larger the connected component has to be. Here SBM and ER match very
    nicely. An example realization is shown in Fig.~\ref{fig:ex_c}.
    The region $O$ does not smoothly go over into region $M$, and both
    coexist for the same size of the giant component, such that both marginal
    probabilities show the two-peak structure, we observed before.

    We will use the end of the results section to make some educated guesses about
    the behavior of planted partitions with more blocks $B$. We would expect that
    basically the same patterns should occur as in the previous case, where the connected
    and biconnected components span different combinations of clusters. Since each cluster
    will have roughly a size of order $1/B$ and the biconnected component may never span
    more clusters than the connected component, there are $B(B+1)/2$ possible combinations,
    which should all manifest as a local maximum
    in the joint probability. Using the same arguments
    as in the $B=2$, we would expect that every local maximum,
    except the one where all clusters
    are part of the biconnected component, corresponding to region M, should be
    exponentially suppressed.

    For the marginal probabilities $P(S)$ (and $P(S_2)$), this suggests that there should
    be $B$ peaks, corresponding to one to $B$ (doubly) connected clusters.
    Like we observed for the $B=2$ case, all but the one corresponding to region M, where all clusters
    are (doubly) connected should be exponentially suppressed.
    However, the width of the peaks in our data suggests that those $B$ peaks might not
    be clearly distinguishable, especially for larger values of $B$.

    \section{Conclusions}
    Here, we studied the distributions of the relative size of the largest
    connected $S$ and biconnected components $S_2$ for the stochastic block model
    with two blocks and strong intra-block connectivity.
    By using sophisticated large-deviation algorithms, we are able to study
    the distributions down to probabilities as small as $10^{-800}$ or below,
    which gives us access to (almost) the full distributions.
    Due to the fast
    convergence to a limiting shape of the empirical rate functions we
    conjecture that the large-deviation principle holds for these distributions.
    Further, we showed where there are
    similarities to the Erd\H{o}s-R\'{e}nyi graph ensemble and for which
    parameters there are differences in different parts of their distributions.
    Especially, we show large qualitative differences in the tails of extremely rare
    events, where the peak regions are indistinguishable.
    These differences seem to be correlated with the
    threshold of the percolation transition. By analyzing the
    correlations between largest connected and largest biconnected component,
    which was also possible in the regime of rare events, we could identify
    three regimes of behavior (cf. Fig.~\ref{fig:scatter}).
    This case study lead to insight into the structure of some rare configurations
    of the SBM, which lead to some educated guesses which behavior is expected
    in other regions of the parameter space.

    In general, our study shows that by analyzing the tails of probability
    distributions for random graphs, differences between ensembles
    can be found which are not detectable by standard simple sampling
    simulations. This is true even if, like here, concerning the percolation
    transition behavior a mapping between
    the effective connectivities of two models exists.
    It is kind of surprising that there are subtle differences in the
    extremely simple observables we studied between two similar models, which are
    inaccessible with a conventional analysis. Thus, large-deviation
    simulations offer access to otherwise hidden properties of networks and
    to correlations between network quantities. Here we demonstrated that they can
    be used in an exploratory way to explore even models with multiple free parameters
    ($N$, $c_\mathrm{inter}$, $c_\mathrm{intra}$, $B$) and guide the search for
    parameter ranges with distinct behavior by allowing the direct examination of
    fringe cases.
    Due to the existence of many different network ensembles, network processes
    and measurable quantities, many new results will likely emerge from applying
    this and similar approaches to gain deep insight into the properties of networks.
    Of particular interest could be to study modified ER models with varying degree
    of (general) (dis-) assortativity by introducing degree correlations,
    to see how this shows up in the distributions
    $P(S)$ and $P(S_2)$, and to compare with the SBM, which is a special
    case. A thorough (but necessarily numerically demanding)
    study could indicate what is a minimal degree of (dis-) assortativity
    to make a difference to the ER significant.

    \section*{Acknowledgments}
        The authors thank Stefan Adolf for performing preliminary studies
        on this topic. We thank Tiago Peixoto for interesting discussions.
        Also, the authors want to thank the two anonymous referees for
        their exceptionally detailed and useful reviews.
        HS acknowledges financial support of the grant HA 3169/8-1 by the
        German Science Foundation (DFG) and the OpLaDyn grant obtained in
        the 4th round of the Trans-Atlantic Platform Digging into Data Challenge
        (2016-147 ANR OPLADYN TAP-DD2016).
        The simulations were performed at the HPC Cluster CARL, located at
        the University of Oldenburg (Germany) and funded by the DFG through
        its Major Research Instrumentation Programme (INST 184/108-1 FUGG)
        and the Ministry of Science and Culture (MWK) of the Lower Saxony
        State.

    \bibliography{lit}

\begin{thebibliography}{43}%
\makeatletter
\providecommand \@ifxundefined [1]{%
 \@ifx{#1\undefined}
}%
\providecommand \@ifnum [1]{%
 \ifnum #1\expandafter \@firstoftwo
 \else \expandafter \@secondoftwo
 \fi
}%
\providecommand \@ifx [1]{%
 \ifx #1\expandafter \@firstoftwo
 \else \expandafter \@secondoftwo
 \fi
}%
\providecommand \natexlab [1]{#1}%
\providecommand \enquote  [1]{``#1''}%
\providecommand \bibnamefont  [1]{#1}%
\providecommand \bibfnamefont [1]{#1}%
\providecommand \citenamefont [1]{#1}%
\providecommand \href@noop [0]{\@secondoftwo}%
\providecommand \href [0]{\begingroup \@sanitize@url \@href}%
\providecommand \@href[1]{\@@startlink{#1}\@@href}%
\providecommand \@@href[1]{\endgroup#1\@@endlink}%
\providecommand \@sanitize@url [0]{\catcode `\\12\catcode `\$12\catcode
  `\&12\catcode `\#12\catcode `\^12\catcode `\_12\catcode `\%12\relax}%
\providecommand \@@startlink[1]{}%
\providecommand \@@endlink[0]{}%
\providecommand \url  [0]{\begingroup\@sanitize@url \@url }%
\providecommand \@url [1]{\endgroup\@href {#1}{\urlprefix }}%
\providecommand \urlprefix  [0]{URL }%
\providecommand \Eprint [0]{\href }%
\providecommand \doibase [0]{https://doi.org/}%
\providecommand \selectlanguage [0]{\@gobble}%
\providecommand \bibinfo  [0]{\@secondoftwo}%
\providecommand \bibfield  [0]{\@secondoftwo}%
\providecommand \translation [1]{[#1]}%
\providecommand \BibitemOpen [0]{}%
\providecommand \bibitemStop [0]{}%
\providecommand \bibitemNoStop [0]{.\EOS\space}%
\providecommand \EOS [0]{\spacefactor3000\relax}%
\providecommand \BibitemShut  [1]{\csname bibitem#1\endcsname}%
\let\auto@bib@innerbib\@empty
\bibitem [{\citenamefont {Holland}\ \emph {et~al.}(1983)\citenamefont
  {Holland}, \citenamefont {Laskey},\ and\ \citenamefont
  {Leinhardt}}]{holland1983stochastic}%
  \BibitemOpen
  \bibfield  {author} {\bibinfo {author} {\bibfnamefont {P.~W.}\ \bibnamefont
  {Holland}}, \bibinfo {author} {\bibfnamefont {K.~B.}\ \bibnamefont
  {Laskey}},\ and\ \bibinfo {author} {\bibfnamefont {S.}~\bibnamefont
  {Leinhardt}},\ }\bibfield  {title} {\bibinfo {title} {Stochastic blockmodels:
  First steps},\ }\href
  {https://doi.org/https://doi.org/10.1016/0378-8733(83)90021-7} {\bibfield
  {journal} {\bibinfo  {journal} {Social Networks}\ }\textbf {\bibinfo {volume}
  {5}},\ \bibinfo {pages} {109 } (\bibinfo {year} {1983})}\BibitemShut
  {NoStop}%
\bibitem [{\citenamefont {Erd\H{o}s}\ and\ \citenamefont
  {R\'enyi}(1960)}]{erdoes1960}%
  \BibitemOpen
  \bibfield  {author} {\bibinfo {author} {\bibfnamefont {P.}~\bibnamefont
  {Erd\H{o}s}}\ and\ \bibinfo {author} {\bibfnamefont {A.}~\bibnamefont
  {R\'enyi}},\ }\bibfield  {title} {\bibinfo {title} {On the evolution of
  random graphs},\ }\href@noop {} {\bibfield  {journal} {\bibinfo  {journal}
  {Publ. Math. Inst. Hungar. Acad. Sci.}\ }\textbf {\bibinfo {volume} {5}},\
  \bibinfo {pages} {17} (\bibinfo {year} {1960})}\BibitemShut {NoStop}%
\bibitem [{\citenamefont {Newman}(2010)}]{newman_book2010}%
  \BibitemOpen
  \bibfield  {author} {\bibinfo {author} {\bibfnamefont {M.}~\bibnamefont
  {Newman}},\ }\href@noop {} {\emph {\bibinfo {title} {Networks: an
  Introduction}}}\ (\bibinfo  {publisher} {Oxford University Press},\ \bibinfo
  {year} {2010})\BibitemShut {NoStop}%
\bibitem [{\citenamefont {Decelle}\ \emph
  {et~al.}(2011{\natexlab{a}})\citenamefont {Decelle}, \citenamefont
  {Krzakala}, \citenamefont {Moore},\ and\ \citenamefont
  {Zdeborov\'a}}]{decelle2011asymptotic}%
  \BibitemOpen
  \bibfield  {author} {\bibinfo {author} {\bibfnamefont {A.}~\bibnamefont
  {Decelle}}, \bibinfo {author} {\bibfnamefont {F.}~\bibnamefont {Krzakala}},
  \bibinfo {author} {\bibfnamefont {C.}~\bibnamefont {Moore}},\ and\ \bibinfo
  {author} {\bibfnamefont {L.}~\bibnamefont {Zdeborov\'a}},\ }\bibfield
  {title} {\bibinfo {title} {Asymptotic analysis of the stochastic block model
  for modular networks and its algorithmic applications},\ }\href
  {https://doi.org/10.1103/PhysRevE.84.066106} {\bibfield  {journal} {\bibinfo
  {journal} {Phys. Rev. E}\ }\textbf {\bibinfo {volume} {84}},\ \bibinfo
  {pages} {066106} (\bibinfo {year} {2011}{\natexlab{a}})}\BibitemShut
  {NoStop}%
\bibitem [{\citenamefont {Karrer}\ and\ \citenamefont
  {Newman}(2011)}]{karrer2011stochastic}%
  \BibitemOpen
  \bibfield  {author} {\bibinfo {author} {\bibfnamefont {B.}~\bibnamefont
  {Karrer}}\ and\ \bibinfo {author} {\bibfnamefont {M.~E.~J.}\ \bibnamefont
  {Newman}},\ }\bibfield  {title} {\bibinfo {title} {Stochastic blockmodels and
  community structure in networks},\ }\href
  {https://doi.org/10.1103/PhysRevE.83.016107} {\bibfield  {journal} {\bibinfo
  {journal} {Phys. Rev. E}\ }\textbf {\bibinfo {volume} {83}},\ \bibinfo
  {pages} {016107} (\bibinfo {year} {2011})}\BibitemShut {NoStop}%
\bibitem [{\citenamefont {Peixoto}(2012)}]{peixoto2012entropy}%
  \BibitemOpen
  \bibfield  {author} {\bibinfo {author} {\bibfnamefont {T.~P.}\ \bibnamefont
  {Peixoto}},\ }\bibfield  {title} {\bibinfo {title} {Entropy of stochastic
  blockmodel ensembles},\ }\href {https://doi.org/10.1103/PhysRevE.85.056122}
  {\bibfield  {journal} {\bibinfo  {journal} {Phys. Rev. E}\ }\textbf {\bibinfo
  {volume} {85}},\ \bibinfo {pages} {056122} (\bibinfo {year}
  {2012})}\BibitemShut {NoStop}%
\bibitem [{\citenamefont {Krzakala}\ \emph {et~al.}(2013)\citenamefont
  {Krzakala}, \citenamefont {Moore}, \citenamefont {Mossel}, \citenamefont
  {Neeman}, \citenamefont {Sly}, \citenamefont {Zdeborov{\'a}},\ and\
  \citenamefont {Zhang}}]{Krzakala2013spectral}%
  \BibitemOpen
  \bibfield  {author} {\bibinfo {author} {\bibfnamefont {F.}~\bibnamefont
  {Krzakala}}, \bibinfo {author} {\bibfnamefont {C.}~\bibnamefont {Moore}},
  \bibinfo {author} {\bibfnamefont {E.}~\bibnamefont {Mossel}}, \bibinfo
  {author} {\bibfnamefont {J.}~\bibnamefont {Neeman}}, \bibinfo {author}
  {\bibfnamefont {A.}~\bibnamefont {Sly}}, \bibinfo {author} {\bibfnamefont
  {L.}~\bibnamefont {Zdeborov{\'a}}},\ and\ \bibinfo {author} {\bibfnamefont
  {P.}~\bibnamefont {Zhang}},\ }\bibfield  {title} {\bibinfo {title} {Spectral
  redemption in clustering sparse networks},\ }\href
  {https://doi.org/10.1073/pnas.1312486110} {\bibfield  {journal} {\bibinfo
  {journal} {Proceedings of the National Academy of Sciences}\ }\textbf
  {\bibinfo {volume} {110}},\ \bibinfo {pages} {20935} (\bibinfo {year}
  {2013})}\BibitemShut {NoStop}%
\bibitem [{\citenamefont {Peixoto}(2014)}]{peixoto2014efficient}%
  \BibitemOpen
  \bibfield  {author} {\bibinfo {author} {\bibfnamefont {T.~P.}\ \bibnamefont
  {Peixoto}},\ }\bibfield  {title} {\bibinfo {title} {Efficient monte carlo and
  greedy heuristic for the inference of stochastic block models},\ }\href
  {https://doi.org/10.1103/PhysRevE.89.012804} {\bibfield  {journal} {\bibinfo
  {journal} {Phys. Rev. E}\ }\textbf {\bibinfo {volume} {89}},\ \bibinfo
  {pages} {012804} (\bibinfo {year} {2014})}\BibitemShut {NoStop}%
\bibitem [{\citenamefont {Peixoto}(2017)}]{peixoto2017nonparametric}%
  \BibitemOpen
  \bibfield  {author} {\bibinfo {author} {\bibfnamefont {T.~P.}\ \bibnamefont
  {Peixoto}},\ }\bibfield  {title} {\bibinfo {title} {Nonparametric bayesian
  inference of the microcanonical stochastic block model},\ }\href
  {https://doi.org/10.1103/PhysRevE.95.012317} {\bibfield  {journal} {\bibinfo
  {journal} {Phys. Rev. E}\ }\textbf {\bibinfo {volume} {95}},\ \bibinfo
  {pages} {012317} (\bibinfo {year} {2017})}\BibitemShut {NoStop}%
\bibitem [{\citenamefont {Decelle}\ \emph
  {et~al.}(2011{\natexlab{b}})\citenamefont {Decelle}, \citenamefont
  {Krzakala}, \citenamefont {Moore},\ and\ \citenamefont
  {Zdeborov\'a}}]{decelle2011inference}%
  \BibitemOpen
  \bibfield  {author} {\bibinfo {author} {\bibfnamefont {A.}~\bibnamefont
  {Decelle}}, \bibinfo {author} {\bibfnamefont {F.}~\bibnamefont {Krzakala}},
  \bibinfo {author} {\bibfnamefont {C.}~\bibnamefont {Moore}},\ and\ \bibinfo
  {author} {\bibfnamefont {L.}~\bibnamefont {Zdeborov\'a}},\ }\bibfield
  {title} {\bibinfo {title} {Inference and phase transitions in the detection
  of modules in sparse networks},\ }\href
  {https://doi.org/10.1103/PhysRevLett.107.065701} {\bibfield  {journal}
  {\bibinfo  {journal} {Phys. Rev. Lett.}\ }\textbf {\bibinfo {volume} {107}},\
  \bibinfo {pages} {065701} (\bibinfo {year} {2011}{\natexlab{b}})}\BibitemShut
  {NoStop}%
\bibitem [{\citenamefont {Nadakuditi}\ and\ \citenamefont
  {Newman}(2012)}]{Nadakuditi2012}%
  \BibitemOpen
  \bibfield  {author} {\bibinfo {author} {\bibfnamefont {R.~R.}\ \bibnamefont
  {Nadakuditi}}\ and\ \bibinfo {author} {\bibfnamefont {M.~E.~J.}\ \bibnamefont
  {Newman}},\ }\bibfield  {title} {\bibinfo {title} {Graph spectra and the
  detectability of community structure in networks},\ }\href
  {https://doi.org/10.1103/PhysRevLett.108.188701} {\bibfield  {journal}
  {\bibinfo  {journal} {Phys. Rev. Lett.}\ }\textbf {\bibinfo {volume} {108}},\
  \bibinfo {pages} {188701} (\bibinfo {year} {2012})}\BibitemShut {NoStop}%
\bibitem [{\citenamefont {Darst}\ \emph {et~al.}(2014)\citenamefont {Darst},
  \citenamefont {Reichman}, \citenamefont {Ronhovde},\ and\ \citenamefont
  {Nussinov}}]{darst2014algorithm}%
  \BibitemOpen
  \bibfield  {author} {\bibinfo {author} {\bibfnamefont {R.~K.}\ \bibnamefont
  {Darst}}, \bibinfo {author} {\bibfnamefont {D.~R.}\ \bibnamefont {Reichman}},
  \bibinfo {author} {\bibfnamefont {P.}~\bibnamefont {Ronhovde}},\ and\
  \bibinfo {author} {\bibfnamefont {Z.}~\bibnamefont {Nussinov}},\ }\bibfield
  {title} {\bibinfo {title} {{Algorithm independent bounds on community
  detection problems and associated transitions in stochastic block model
  graphs}},\ }\href {https://doi.org/10.1093/comnet/cnu042} {\bibfield
  {journal} {\bibinfo  {journal} {Journal of Complex Networks}\ }\textbf
  {\bibinfo {volume} {3}},\ \bibinfo {pages} {333} (\bibinfo {year}
  {2014})}\BibitemShut {NoStop}%
\bibitem [{\citenamefont {Albert}\ \emph {et~al.}(2000)\citenamefont {Albert},
  \citenamefont {Jeong},\ and\ \citenamefont {Barab{\'a}si}}]{albert2000error}%
  \BibitemOpen
  \bibfield  {author} {\bibinfo {author} {\bibfnamefont {R.}~\bibnamefont
  {Albert}}, \bibinfo {author} {\bibfnamefont {H.}~\bibnamefont {Jeong}},\ and\
  \bibinfo {author} {\bibfnamefont {A.-L.}\ \bibnamefont {Barab{\'a}si}},\
  }\bibfield  {title} {\bibinfo {title} {Error and attack tolerance of complex
  networks},\ }\href {https://doi.org/10.1038/35019019} {\bibfield  {journal}
  {\bibinfo  {journal} {Nature}\ }\textbf {\bibinfo {volume} {406}},\ \bibinfo
  {pages} {378} (\bibinfo {year} {2000})}\BibitemShut {NoStop}%
\bibitem [{\citenamefont {Norrenbrock}\ \emph {et~al.}(2016)\citenamefont
  {Norrenbrock}, \citenamefont {Melchert},\ and\ \citenamefont
  {Hartmann}}]{Norrenbrock2016fragmentation}%
  \BibitemOpen
  \bibfield  {author} {\bibinfo {author} {\bibfnamefont {C.}~\bibnamefont
  {Norrenbrock}}, \bibinfo {author} {\bibfnamefont {O.}~\bibnamefont
  {Melchert}},\ and\ \bibinfo {author} {\bibfnamefont {A.~K.}\ \bibnamefont
  {Hartmann}},\ }\bibfield  {title} {\bibinfo {title} {Fragmentation properties
  of two-dimensional proximity graphs considering random failures and targeted
  attacks},\ }\href {https://doi.org/10.1103/PhysRevE.94.062125} {\bibfield
  {journal} {\bibinfo  {journal} {Phys. Rev. E}\ }\textbf {\bibinfo {volume}
  {94}},\ \bibinfo {pages} {062125} (\bibinfo {year} {2016})}\BibitemShut
  {NoStop}%
\bibitem [{\citenamefont {Callaway}\ \emph {et~al.}(2000)\citenamefont
  {Callaway}, \citenamefont {Newman}, \citenamefont {Strogatz},\ and\
  \citenamefont {Watts}}]{Callaway2000network}%
  \BibitemOpen
  \bibfield  {author} {\bibinfo {author} {\bibfnamefont {D.~S.}\ \bibnamefont
  {Callaway}}, \bibinfo {author} {\bibfnamefont {M.~E.~J.}\ \bibnamefont
  {Newman}}, \bibinfo {author} {\bibfnamefont {S.~H.}\ \bibnamefont
  {Strogatz}},\ and\ \bibinfo {author} {\bibfnamefont {D.~J.}\ \bibnamefont
  {Watts}},\ }\bibfield  {title} {\bibinfo {title} {Network robustness and
  fragility: Percolation on random graphs},\ }\href
  {https://doi.org/10.1103/PhysRevLett.85.5468} {\bibfield  {journal} {\bibinfo
   {journal} {Phys. Rev. Lett.}\ }\textbf {\bibinfo {volume} {85}},\ \bibinfo
  {pages} {5468} (\bibinfo {year} {2000})}\BibitemShut {NoStop}%
\bibitem [{\citenamefont {Cohen}\ \emph {et~al.}(2000)\citenamefont {Cohen},
  \citenamefont {Erez}, \citenamefont {ben Avraham},\ and\ \citenamefont
  {Havlin}}]{Cohen2000Resilience}%
  \BibitemOpen
  \bibfield  {author} {\bibinfo {author} {\bibfnamefont {R.}~\bibnamefont
  {Cohen}}, \bibinfo {author} {\bibfnamefont {K.}~\bibnamefont {Erez}},
  \bibinfo {author} {\bibfnamefont {D.}~\bibnamefont {ben Avraham}},\ and\
  \bibinfo {author} {\bibfnamefont {S.}~\bibnamefont {Havlin}},\ }\bibfield
  {title} {\bibinfo {title} {Resilience of the internet to random breakdowns},\
  }\href {https://doi.org/10.1103/PhysRevLett.85.4626} {\bibfield  {journal}
  {\bibinfo  {journal} {Phys. Rev. Lett.}\ }\textbf {\bibinfo {volume} {85}},\
  \bibinfo {pages} {4626} (\bibinfo {year} {2000})}\BibitemShut {NoStop}%
\bibitem [{\citenamefont {Dewenter}\ and\ \citenamefont
  {Hartmann}(2015)}]{dewenter2015large}%
  \BibitemOpen
  \bibfield  {author} {\bibinfo {author} {\bibfnamefont {T.}~\bibnamefont
  {Dewenter}}\ and\ \bibinfo {author} {\bibfnamefont {A.~K.}\ \bibnamefont
  {Hartmann}},\ }\bibfield  {title} {\bibinfo {title} {Large-deviation
  properties of resilience of power grids},\ }\href
  {http://stacks.iop.org/1367-2630/17/i=1/a=015005} {\bibfield  {journal}
  {\bibinfo  {journal} {New Journal of Physics}\ }\textbf {\bibinfo {volume}
  {17}},\ \bibinfo {pages} {015005} (\bibinfo {year} {2015})}\BibitemShut
  {NoStop}%
\bibitem [{\citenamefont {Newman}\ and\ \citenamefont
  {Ghoshal}(2008)}]{Newman2008Bicomponents}%
  \BibitemOpen
  \bibfield  {author} {\bibinfo {author} {\bibfnamefont {M.~E.~J.}\
  \bibnamefont {Newman}}\ and\ \bibinfo {author} {\bibfnamefont
  {G.}~\bibnamefont {Ghoshal}},\ }\bibfield  {title} {\bibinfo {title}
  {Bicomponents and the robustness of networks to failure},\ }\href
  {https://doi.org/10.1103/PhysRevLett.100.138701} {\bibfield  {journal}
  {\bibinfo  {journal} {Phys. Rev. Lett.}\ }\textbf {\bibinfo {volume} {100}},\
  \bibinfo {pages} {138701} (\bibinfo {year} {2008})}\BibitemShut {NoStop}%
\bibitem [{\citenamefont {Touchette}(2009)}]{touchette2009}%
  \BibitemOpen
  \bibfield  {author} {\bibinfo {author} {\bibfnamefont {H.}~\bibnamefont
  {Touchette}},\ }\bibfield  {title} {\bibinfo {title} {The large deviation
  approach to statistical mechanics},\ }\href
  {https://doi.org/10.1016/j.physrep.2009.05.002} {\bibfield  {journal}
  {\bibinfo  {journal} {Physics Reports}\ }\textbf {\bibinfo {volume} {478}},\
  \bibinfo {pages} {1 } (\bibinfo {year} {2009})}\BibitemShut {NoStop}%
\bibitem [{\citenamefont {Biskup}\ \emph {et~al.}(2007)\citenamefont {Biskup},
  \citenamefont {Chayes},\ and\ \citenamefont {Smith}}]{biskup2007large}%
  \BibitemOpen
  \bibfield  {author} {\bibinfo {author} {\bibfnamefont {M.}~\bibnamefont
  {Biskup}}, \bibinfo {author} {\bibfnamefont {L.}~\bibnamefont {Chayes}},\
  and\ \bibinfo {author} {\bibfnamefont {S.~A.}\ \bibnamefont {Smith}},\
  }\bibfield  {title} {\bibinfo {title} {Large-deviations/thermodynamic
  approach to percolation on the complete graph},\ }\href@noop {} {\bibfield
  {journal} {\bibinfo  {journal} {Random Structures \& Algorithms}\ }\textbf
  {\bibinfo {volume} {31}},\ \bibinfo {pages} {354} (\bibinfo {year}
  {2007})}\BibitemShut {NoStop}%
\bibitem [{\citenamefont {Hartmann}(2011)}]{Hartmann2011large}%
  \BibitemOpen
  \bibfield  {author} {\bibinfo {author} {\bibfnamefont {A.~K.}\ \bibnamefont
  {Hartmann}},\ }\bibfield  {title} {\bibinfo {title} {Large-deviation
  properties of largest component for random graphs},\ }\href
  {https://doi.org/10.1140/epjb/e2011-10836-4} {\bibfield  {journal} {\bibinfo
  {journal} {The European Physical Journal B}\ }\textbf {\bibinfo {volume}
  {84}},\ \bibinfo {pages} {627} (\bibinfo {year} {2011})}\BibitemShut
  {NoStop}%
\bibitem [{\citenamefont {Schawe}\ and\ \citenamefont
  {Hartmann}(2019)}]{Schawe2019large}%
  \BibitemOpen
  \bibfield  {author} {\bibinfo {author} {\bibfnamefont {H.}~\bibnamefont
  {Schawe}}\ and\ \bibinfo {author} {\bibfnamefont {A.~K.}\ \bibnamefont
  {Hartmann}},\ }\bibfield  {title} {\bibinfo {title} {Large-deviation
  properties of the largest biconnected component for random graphs},\ }\href
  {https://doi.org/10.1140/epjb/e2019-90667-y} {\bibfield  {journal} {\bibinfo
  {journal} {The European Physical Journal B}\ }\textbf {\bibinfo {volume}
  {92}},\ \bibinfo {pages} {73} (\bibinfo {year} {2019})}\BibitemShut {NoStop}%
\bibitem [{\citenamefont {Hartmann}\ and\ \citenamefont
  {M\'{e}zard}(2018)}]{hartmann2018distribution}%
  \BibitemOpen
  \bibfield  {author} {\bibinfo {author} {\bibfnamefont {A.~K.}\ \bibnamefont
  {Hartmann}}\ and\ \bibinfo {author} {\bibfnamefont {M.}~\bibnamefont
  {M\'{e}zard}},\ }\bibfield  {title} {\bibinfo {title} {Distribution of
  diameters for {Erd\H{o}s-R\'{e}nyi} random graphs},\ }\href
  {https://doi.org/10.1103/PhysRevE.97.032128} {\bibfield  {journal} {\bibinfo
  {journal} {Phys. Rev. E}\ }\textbf {\bibinfo {volume} {97}},\ \bibinfo
  {pages} {032128} (\bibinfo {year} {2018})}\BibitemShut {NoStop}%
\bibitem [{\citenamefont {Hartmann}(2017)}]{Hartmann2017Large}%
  \BibitemOpen
  \bibfield  {author} {\bibinfo {author} {\bibfnamefont {A.~K.}\ \bibnamefont
  {Hartmann}},\ }\bibfield  {title} {\bibinfo {title} {Large-deviation
  properties of the largest 2-core component for random graphs},\ }\href
  {https://doi.org/10.1140/epjst/e2016-60368-3} {\bibfield  {journal} {\bibinfo
   {journal} {The European Physical Journal Special Topics}\ }\textbf {\bibinfo
  {volume} {226}},\ \bibinfo {pages} {567} (\bibinfo {year}
  {2017})}\BibitemShut {NoStop}%
\bibitem [{\citenamefont {Barab{\'a}si}\ and\ \citenamefont
  {Albert}(1999)}]{barabasi1999emergence}%
  \BibitemOpen
  \bibfield  {author} {\bibinfo {author} {\bibfnamefont {A.-L.}\ \bibnamefont
  {Barab{\'a}si}}\ and\ \bibinfo {author} {\bibfnamefont {R.}~\bibnamefont
  {Albert}},\ }\bibfield  {title} {\bibinfo {title} {Emergence of scaling in
  random networks},\ }\href@noop {} {\bibfield  {journal} {\bibinfo  {journal}
  {Science}\ }\textbf {\bibinfo {volume} {286}},\ \bibinfo {pages} {509}
  (\bibinfo {year} {1999})}\BibitemShut {NoStop}%
\bibitem [{\citenamefont {Hopcroft}\ and\ \citenamefont
  {Tarjan}(1973)}]{Hopcroft1973algorithm}%
  \BibitemOpen
  \bibfield  {author} {\bibinfo {author} {\bibfnamefont {J.}~\bibnamefont
  {Hopcroft}}\ and\ \bibinfo {author} {\bibfnamefont {R.}~\bibnamefont
  {Tarjan}},\ }\bibfield  {title} {\bibinfo {title} {Algorithm 447: Efficient
  algorithms for graph manipulation},\ }\href
  {https://doi.org/10.1145/362248.362272} {\bibfield  {journal} {\bibinfo
  {journal} {Commun. ACM}\ }\textbf {\bibinfo {volume} {16}},\ \bibinfo {pages}
  {372} (\bibinfo {year} {1973})}\BibitemShut {NoStop}%
\bibitem [{\citenamefont {Cormen}\ \emph {et~al.}(2009)\citenamefont {Cormen},
  \citenamefont {Leiserson}, \citenamefont {Rivest},\ and\ \citenamefont
  {Stein}}]{cormen2009introduction}%
  \BibitemOpen
  \bibfield  {author} {\bibinfo {author} {\bibfnamefont {T.~H.}\ \bibnamefont
  {Cormen}}, \bibinfo {author} {\bibfnamefont {C.~E.}\ \bibnamefont
  {Leiserson}}, \bibinfo {author} {\bibfnamefont {R.~L.}\ \bibnamefont
  {Rivest}},\ and\ \bibinfo {author} {\bibfnamefont {C.}~\bibnamefont
  {Stein}},\ }\href@noop {} {\emph {\bibinfo {title} {Introduction to
  algorithms}}}\ (\bibinfo  {publisher} {MIT press},\ \bibinfo {year}
  {2009})\BibitemShut {NoStop}%
\bibitem [{\citenamefont {Dezs\H{o}}\ \emph {et~al.}(2011)\citenamefont
  {Dezs\H{o}}, \citenamefont {J{\"u}ttner},\ and\ \citenamefont
  {Kov\'acs}}]{lemon}%
  \BibitemOpen
  \bibfield  {author} {\bibinfo {author} {\bibfnamefont {B.}~\bibnamefont
  {Dezs\H{o}}}, \bibinfo {author} {\bibfnamefont {A.}~\bibnamefont
  {J{\"u}ttner}},\ and\ \bibinfo {author} {\bibfnamefont {P.}~\bibnamefont
  {Kov\'acs}},\ }\bibfield  {title} {\bibinfo {title} {Lemon – an open source
  c++ graph template library},\ }\href
  {https://doi.org/https://doi.org/10.1016/j.entcs.2011.06.003} {\bibfield
  {journal} {\bibinfo  {journal} {Electronic Notes in Theoretical Computer
  Science}\ }\textbf {\bibinfo {volume} {264}},\ \bibinfo {pages} {23 }
  (\bibinfo {year} {2011})},\ \bibinfo {note} {proceedings of the Second
  Workshop on Generative Technologies (WGT) 2010}\BibitemShut {NoStop}%
\bibitem [{\citenamefont {Hartmann}\ and\ \citenamefont
  {Weigt}(2005)}]{hartmann2005phase}%
  \BibitemOpen
  \bibfield  {author} {\bibinfo {author} {\bibfnamefont {A.~K.}\ \bibnamefont
  {Hartmann}}\ and\ \bibinfo {author} {\bibfnamefont {M.}~\bibnamefont
  {Weigt}},\ }\href@noop {} {\emph {\bibinfo {title} {Phase transitions in
  combinatorial optimization problems}}}\ (\bibinfo  {publisher} {Wiley Online
  Library},\ \bibinfo {year} {2005})\BibitemShut {NoStop}%
\bibitem [{\citenamefont {Bryc}(1993)}]{bryc1993remark}%
  \BibitemOpen
  \bibfield  {author} {\bibinfo {author} {\bibfnamefont {W.}~\bibnamefont
  {Bryc}},\ }\bibfield  {title} {\bibinfo {title} {A remark on the connection
  between the large deviation principle and the central limit theorem},\ }\href
  {https://doi.org/10.1016/0167-7152(93)90012-8} {\bibfield  {journal}
  {\bibinfo  {journal} {Statistics \& Probability Letters}\ }\textbf {\bibinfo
  {volume} {18}},\ \bibinfo {pages} {253 } (\bibinfo {year}
  {1993})}\BibitemShut {NoStop}%
\bibitem [{\citenamefont {Hartmann}(2015)}]{practical_guide2015}%
  \BibitemOpen
  \bibfield  {author} {\bibinfo {author} {\bibfnamefont {A.~K.}\ \bibnamefont
  {Hartmann}},\ }\href@noop {} {\emph {\bibinfo {title} {{Big Practical Guide
  to Computer Simulations}}}}\ (\bibinfo  {publisher} {World Scientific},\
  \bibinfo {address} {Singapore},\ \bibinfo {year} {2015})\BibitemShut
  {NoStop}%
\bibitem [{\citenamefont {Wang}\ and\ \citenamefont
  {Landau}(2001{\natexlab{a}})}]{Wang2001Efficient}%
  \BibitemOpen
  \bibfield  {author} {\bibinfo {author} {\bibfnamefont {F.}~\bibnamefont
  {Wang}}\ and\ \bibinfo {author} {\bibfnamefont {D.~P.}\ \bibnamefont
  {Landau}},\ }\bibfield  {title} {\bibinfo {title} {Efficient, multiple-range
  random walk algorithm to calculate the density of states},\ }\href
  {https://doi.org/10.1103/PhysRevLett.86.2050} {\bibfield  {journal} {\bibinfo
   {journal} {Phys. Rev. Lett.}\ }\textbf {\bibinfo {volume} {86}},\ \bibinfo
  {pages} {2050} (\bibinfo {year} {2001}{\natexlab{a}})}\BibitemShut {NoStop}%
\bibitem [{\citenamefont {Wang}\ and\ \citenamefont
  {Landau}(2001{\natexlab{b}})}]{Wang2001Determining}%
  \BibitemOpen
  \bibfield  {author} {\bibinfo {author} {\bibfnamefont {F.}~\bibnamefont
  {Wang}}\ and\ \bibinfo {author} {\bibfnamefont {D.~P.}\ \bibnamefont
  {Landau}},\ }\bibfield  {title} {\bibinfo {title} {Determining the density of
  states for classical statistical models: A random walk algorithm to produce a
  flat histogram},\ }\href {https://doi.org/10.1103/PhysRevE.64.056101}
  {\bibfield  {journal} {\bibinfo  {journal} {Phys. Rev. E}\ }\textbf {\bibinfo
  {volume} {64}},\ \bibinfo {pages} {056101} (\bibinfo {year}
  {2001}{\natexlab{b}})}\BibitemShut {NoStop}%
\bibitem [{\citenamefont {Dickman}\ and\ \citenamefont
  {Cunha-Netto}(2011)}]{Dickman2011Complete}%
  \BibitemOpen
  \bibfield  {author} {\bibinfo {author} {\bibfnamefont {R.}~\bibnamefont
  {Dickman}}\ and\ \bibinfo {author} {\bibfnamefont {A.~G.}\ \bibnamefont
  {Cunha-Netto}},\ }\bibfield  {title} {\bibinfo {title} {Complete
  high-precision entropic sampling},\ }\href
  {https://doi.org/10.1103/PhysRevE.84.026701} {\bibfield  {journal} {\bibinfo
  {journal} {Phys. Rev. E}\ }\textbf {\bibinfo {volume} {84}},\ \bibinfo
  {pages} {026701} (\bibinfo {year} {2011})}\BibitemShut {NoStop}%
\bibitem [{\citenamefont {Lee}(1993)}]{Lee1993Entropic}%
  \BibitemOpen
  \bibfield  {author} {\bibinfo {author} {\bibfnamefont {J.}~\bibnamefont
  {Lee}},\ }\bibfield  {title} {\bibinfo {title} {New {Monte Carlo} algorithm:
  Entropic sampling},\ }\href {https://doi.org/10.1103/PhysRevLett.71.211}
  {\bibfield  {journal} {\bibinfo  {journal} {Phys. Rev. Lett.}\ }\textbf
  {\bibinfo {volume} {71}},\ \bibinfo {pages} {211} (\bibinfo {year}
  {1993})}\BibitemShut {NoStop}%
\bibitem [{\citenamefont {Belardinelli}\ and\ \citenamefont
  {Pereyra}(2007{\natexlab{a}})}]{Belardinelli2007Fast}%
  \BibitemOpen
  \bibfield  {author} {\bibinfo {author} {\bibfnamefont {R.~E.}\ \bibnamefont
  {Belardinelli}}\ and\ \bibinfo {author} {\bibfnamefont {V.~D.}\ \bibnamefont
  {Pereyra}},\ }\bibfield  {title} {\bibinfo {title} {Fast algorithm to
  calculate density of states},\ }\href
  {https://doi.org/10.1103/PhysRevE.75.046701} {\bibfield  {journal} {\bibinfo
  {journal} {Phys. Rev. E}\ }\textbf {\bibinfo {volume} {75}},\ \bibinfo
  {pages} {046701} (\bibinfo {year} {2007}{\natexlab{a}})}\BibitemShut
  {NoStop}%
\bibitem [{\citenamefont {Belardinelli}\ and\ \citenamefont
  {Pereyra}(2007{\natexlab{b}})}]{Belardinelli2007theoretical}%
  \BibitemOpen
  \bibfield  {author} {\bibinfo {author} {\bibfnamefont {R.~E.}\ \bibnamefont
  {Belardinelli}}\ and\ \bibinfo {author} {\bibfnamefont {V.~D.}\ \bibnamefont
  {Pereyra}},\ }\bibfield  {title} {\bibinfo {title} {{Wang-Landau} algorithm:
  A theoretical analysis of the saturation of the error},\ }\href
  {https://doi.org/10.1063/1.2803061} {\bibfield  {journal} {\bibinfo
  {journal} {The Journal of Chemical Physics}\ }\textbf {\bibinfo {volume}
  {127}},\ \bibinfo {eid} {184105} (\bibinfo {year}
  {2007}{\natexlab{b}})}\BibitemShut {NoStop}%
\bibitem [{Note1()}]{Note1}%
  \BibitemOpen
  \bibinfo {note} {Intel Xeon E5-2650 v4}\BibitemShut {NoStop}%
\bibitem [{\citenamefont {Schawe}\ \emph {et~al.}(2018)\citenamefont {Schawe},
  \citenamefont {Hartmann},\ and\ \citenamefont
  {Majumdar}}]{schawe2018avoiding}%
  \BibitemOpen
  \bibfield  {author} {\bibinfo {author} {\bibfnamefont {H.}~\bibnamefont
  {Schawe}}, \bibinfo {author} {\bibfnamefont {A.~K.}\ \bibnamefont
  {Hartmann}},\ and\ \bibinfo {author} {\bibfnamefont {S.~N.}\ \bibnamefont
  {Majumdar}},\ }\bibfield  {title} {\bibinfo {title} {Large deviations of
  convex hulls of self-avoiding random walks},\ }\href
  {https://doi.org/10.1103/PhysRevE.97.062159} {\bibfield  {journal} {\bibinfo
  {journal} {Phys. Rev. E}\ }\textbf {\bibinfo {volume} {97}},\ \bibinfo
  {pages} {062159} (\bibinfo {year} {2018})}\BibitemShut {NoStop}%
\bibitem [{\citenamefont {Epps}\ and\ \citenamefont
  {Singleton}(1986)}]{epps1986omnibus}%
  \BibitemOpen
  \bibfield  {author} {\bibinfo {author} {\bibfnamefont {T.}~\bibnamefont
  {Epps}}\ and\ \bibinfo {author} {\bibfnamefont {K.~J.}\ \bibnamefont
  {Singleton}},\ }\bibfield  {title} {\bibinfo {title} {An omnibus test for the
  two-sample problem using the empirical characteristic function},\ }\href
  {https://doi.org/10.1080/00949658608810963} {\bibfield  {journal} {\bibinfo
  {journal} {Journal of Statistical Computation and Simulation}\ }\textbf
  {\bibinfo {volume} {26}},\ \bibinfo {pages} {177} (\bibinfo {year}
  {1986})}\BibitemShut {NoStop}%
\bibitem [{\citenamefont {Virtanen}\ \emph {et~al.}(2020)\citenamefont
  {Virtanen}, \citenamefont {Gommers}, \citenamefont {Oliphant}, \citenamefont
  {Haberland}, \citenamefont {Reddy}, \citenamefont {Cournapeau}, \citenamefont
  {Burovski}, \citenamefont {Peterson}, \citenamefont {Weckesser},
  \citenamefont {Bright} \emph {et~al.}}]{scipy}%
  \BibitemOpen
  \bibfield  {author} {\bibinfo {author} {\bibfnamefont {P.}~\bibnamefont
  {Virtanen}}, \bibinfo {author} {\bibfnamefont {R.}~\bibnamefont {Gommers}},
  \bibinfo {author} {\bibfnamefont {T.~E.}\ \bibnamefont {Oliphant}}, \bibinfo
  {author} {\bibfnamefont {M.}~\bibnamefont {Haberland}}, \bibinfo {author}
  {\bibfnamefont {T.}~\bibnamefont {Reddy}}, \bibinfo {author} {\bibfnamefont
  {D.}~\bibnamefont {Cournapeau}}, \bibinfo {author} {\bibfnamefont
  {E.}~\bibnamefont {Burovski}}, \bibinfo {author} {\bibfnamefont
  {P.}~\bibnamefont {Peterson}}, \bibinfo {author} {\bibfnamefont
  {W.}~\bibnamefont {Weckesser}}, \bibinfo {author} {\bibfnamefont
  {J.}~\bibnamefont {Bright}}, \emph {et~al.},\ }\bibfield  {title} {\bibinfo
  {title} {Scipy 1.0: fundamental algorithms for scientific computing in
  python},\ }\href@noop {} {\bibfield  {journal} {\bibinfo  {journal} {Nature
  methods}\ }\textbf {\bibinfo {volume} {17}},\ \bibinfo {pages} {261}
  (\bibinfo {year} {2020})}\BibitemShut {NoStop}%
\bibitem [{\citenamefont {Press}\ \emph {et~al.}(2007)\citenamefont {Press},
  \citenamefont {Teukolsky}, \citenamefont {Vetterling},\ and\ \citenamefont
  {Flannery}}]{press2007numerical}%
  \BibitemOpen
  \bibfield  {author} {\bibinfo {author} {\bibfnamefont {W.~H.}\ \bibnamefont
  {Press}}, \bibinfo {author} {\bibfnamefont {S.~A.}\ \bibnamefont
  {Teukolsky}}, \bibinfo {author} {\bibfnamefont {W.~T.}\ \bibnamefont
  {Vetterling}},\ and\ \bibinfo {author} {\bibfnamefont {B.~P.}\ \bibnamefont
  {Flannery}},\ }\href@noop {} {\emph {\bibinfo {title} {Numerical recipes 3rd
  edition: The art of scientific computing}}}\ (\bibinfo  {publisher}
  {Cambridge university press},\ \bibinfo {year} {2007})\BibitemShut {NoStop}%
\bibitem [{\citenamefont {Anderson}\ and\ \citenamefont
  {Darling}(1952)}]{anderson1952asymptotic}%
  \BibitemOpen
  \bibfield  {author} {\bibinfo {author} {\bibfnamefont {T.~W.}\ \bibnamefont
  {Anderson}}\ and\ \bibinfo {author} {\bibfnamefont {D.~A.}\ \bibnamefont
  {Darling}},\ }\bibfield  {title} {\bibinfo {title} {Asymptotic theory of
  certain "goodness of fit" criteria based on stochastic processes},\ }\href
  {http://www.jstor.org/stable/2236446} {\bibfield  {journal} {\bibinfo
  {journal} {The Annals of Mathematical Statistics}\ }\textbf {\bibinfo
  {volume} {23}},\ \bibinfo {pages} {193} (\bibinfo {year} {1952})}\BibitemShut
  {NoStop}%
\end{thebibliography}%

\end{document}